\documentclass{article}

\usepackage{amscd,amsmath,amsthm,amssymb,latexsym,cite,finalT}

\input eym_gb.def

\newcommand{\sauthor}{T. Oliynyk and H.P. K\"unzle}
\newcommand{\stitle}{Global behavior of EYM solutions}

\numberwithin{equation}{section}

\begin{document}

\title{Global behavior of solutions to the static spherically symmetric
EYM equations 
\thanks{PACS: 04.40.Nr, 11.15.Kc}
\thanks{2000 \emph{Mathematics Subject Classification} 
Primary 83C20, 83C22; Secondary 53C30, 17B81.}
}
\author{
Todd A. Oliynyk \thanks{todd.oliynyk@ise.canberra.edu.au . Present address: 
School of Mathematics and Statistics, University of Canberra,
ACT 2601, Australia. }\\[3mm]
H.P. K\"{u}nzle \thanks{hp.kunzle@ualberta.ca}\\[3mm]
Department of Mathematical and Statistical Sciences\\
University of Alberta\\
Edmonton, Canada T6G 2G1}

\date{}
\maketitle

\begin{abstract}
  
The set of all possible spherically symmetric magnetic static
Einstein-Yang-Mills field equations for an arbitrary compact semi-simple
gauge group $G$ was classified in two previous papers. Local analytic
solutions near the center and a black hole horizon as well as those that are
analytic and bounded near infinity were shown to exist. Some globally
bounded solutions are also known to exist because they can be obtained by
embedding solutions for the $G=SU(2)$ case which is well understood. Here we
derive some asymptotic properties of an arbitrary global solution, namely
one that exists locally near a radial value $r_{0}$, has positive mass
$m(r)$ at $r_{0}$ and develops no horizon for all $r>r_{0}$. The set of 
asymptotic values of the Yang-Mills potential (in a suitable well 
defined gauge) is shown to be finite in the so-called regular case, but 
may form a more complicated real variety for models obtained from 
irregular rotation group actions.

\end{abstract}

\sect{chintro}{Introduction}

Static spherically symmetric and globally regular solutions to the
Einstein-Yang-Mills (EYM) equations for the gauge group $SU(2)$ were first
found numerically by Bartnik and Mckinnon \cite{k4752}. They showed that
although there are no static, finite energy pure Yang-Mills fields
\cite{k4861},\cite{k5272} when coupled to the gravitational instead of the
Higgs field physically meaningfull classical solutions where possible.  As is
standard, we will refer to these static, globally regular solutions as
\emph{solitons}. Shortly after the solitons were discovered, static
spherically symmetric $SU(2)$-EYM black holes were found numerically
\cite{hka25,k5062,k4971}.

The $SU(2)$-EYM black hole solutions provided counter examples to the no-hair
conjecture and stimulated investigations into other matter fields coupled to
gravity for the purpose of finding solutions that violated the no hair
conjecture.  Consequently, it has been realized that violation of the no hair
conjecture is typical for gravity coupled to non-Abelian gauge theories.  More
recently \cite{k6190,k6298}, axisymmetric static black hole and soliton
solutions to the $SU(2)$-EYM equations have been constructed numerically,
providing dramatic examples of violations to the no hair conjecture.  All of
these solutions have shown that equilibrium configurations of black holes can
be much more complicated than had been previously thought.

Existence of the soliton and black hole solutions to the $SU(2)$-EYM equations
was first established analytically by Smoller, Wasserman, Yau, and McLeod
\cite{k5061,k5185,k5201}.  Global existence was also established by
Breitenlohner, Forg\'acs, and Maison in \cite{k5428} using different methods.
Smoller and Wasserman have extensively studied the $SU(2)$-EYM equations
\cite{k5516,k5719,k5824,k5825,k6135,k6377} and have completely classified
\cite{k6560} the solutions which are defined in the far field, i.e. for large
radius $r$.  One surprising result that they have discovered is that any
solution that is defined in the far field is actually defined on the whole
interval $(0,\infty)$. This is not the usual situation for solutions to
non-linear systems of differential equations where one normally expects global
existence for only a small subset of the initial conditions. On the other hand
for such solutions the mass will typically be negative below some limiting
radius, and there is a singularity at r=0.

For gauge groups $G$ other than $SU(2)$, much less is known and the
investigations have almost exclusively focused on $SU(n)$ and only for the
most obvious ansatz for the spherically symmetric gauge field.  Recently, the
existence of $SU(3)$ black hole solutions which are not embedded $SU(2)$
solutions has been established analytically \cite{k6645}.  For $SU(3)$ and
$SU(4)$ both solitons and black hole solutions which are not embedded $SU(2)$
solutions have been found numerically \cite{hka28,k5749,k6485}. From the
numerical solutions it is clear that the existence proof in \cite{k6645} does
not cover all the possibilities. So even in $SU(3)$ there is still work to be
done.

For arbitrary compact gauge groups even less is known. No numerical solutions
have been constructed for $G\neq SU(n)$ and the only analytical work we are
aware of is contained in the papers \cite{k5109,k4295,k5626}.  However, in
these papers, only the so called \emph{regular actions} of the symmetry group
are considered which, as we shall see later, is a strong condition. For a
review of these developments in EYM theory, see \cite{k6373}. The EYM
equations continue to attract attention. Rotating $SU(2)$-EYM black holes have
been constructed numerically \cite{k6546} and the $SU(2)$-EYM equations with a
cosmological constant have been studied
\cite{k6499,k6500,k6501,k6388,k6620,k6642}.

In this paper we show that a number of the analytical results derived in
\cite{hka25,k5061,k5185,k5201,k5516,k5719,k5824,k5825,k6135,k6377} generalize
to ``magnetic type'' EYM solutions for arbitrary compact gauge groups and
arbitrary actions of the symmetry group $K=SU(2)$. More precisely, our main
result (Theorem \ref{gbthm21}) is the behavior for $r\ra\infty$ of the
solution under the assumption that it exists at some value $r_{0}$ and that
the mass is positive at $r_{0}$ and no horizon develops for $r>r_{0}$.  
It follows
easily from a scaling argument that any global solution of the $SU(2)$ theory
induces ``embedded solutions'' for any model with arbitrary $G$ and arbitrary
action of $K$. However, these special solutions are extremely unstable against
the slightest change of the parameters that determine it near $0$ and
$\infty$. In fact, numerical experimentation shows that the same probably
holds for most other solutions, if any, which makes a general numerical
exploration of the set of all global solutions extremely difficult if not
almost impossible. This unfortunate situation is not quite unexpected in view
of the instability results with respect to time dependence obtained in
\cite{k5626} for regular actions and general compact gauge groups.

The article is organized as follows. Sections \ref{liealg} and \ref{sl2}
contain a review of the results from Lie algebra theory and the theory of
three dimensional semisimple Lie algebras based largely on \cite{k5157} and
\cite{k6494}.  The static spherically symmetric EYM equations are presented in
section \ref{eymeqns}.  In sections \ref{restLo} and \ref{regA1} we include a
determination of which actions are \emph{regular} for the classical simple
gauge groups based on the orbit theory of \cite{k6494}. (We have previously
computed them for the exceptional Lie algebras in \cite{eymgg}.)  Our main
result concerning the asymptotic behavior of solutions to the EYM equations
(Theorem \ref{gbthm21}) is given in section \ref{chgb}. The remainder of the
paper is devoted to the proof of this theorem.

\sect{liealg}{Preliminaries}

In this section we fix our Lie algebra notation and
collect some results from
Lie algebra theory that will be required in the following
sections.  Many of the results are well known and
can be found, for example, in
\cite{k5157} and \cite{k6082}.
Throughout this article $G$ will always denote a real compact semisimple
Lie group with Lie algebra $\g_{0}$. 
The adjoint action of $G$ on $\g_{0}$ will be 
denoted by $\Ad$, while $\ad$ will denote the
adjoint action of $\g_{0}$ on $\g_{0}$, i.e. 
$\ad(X)(Y) = [X,Y]$ for all $X$,$Y$ in $\g_{0}$.
The complexification of $\g_{0}$ will be denoted by $\g$.
The $\ad$ action can then
be extended by complex linearity to an action of $\g$ on
$\g$ so that $\ad(X)(Y) := [X,Y]$ for all $X$,$Y$ in $\g$.
We will use the notation $\g^{X}$ for the centralizer of
a single element $X\in \g$. In other words,
\eqn{gX}{
\g^{X} := \{\: Y\in \g\:|\: [X,Y]=0 \:\} \: .
}
Similarly, $\g_{0}^{X}$ is the centralizer of
an element $X\in \g_{0}$.

We will let $\Kill{\cdot}{\cdot}$ be any non-degenerate
$\ad$-invariant bilinear form on $\g$ that restricts to
a negative definite inner product on $\g_{0}$. By $\ad$-invariance
we mean that
\eqn{adinvar}{
\Kill{[X,Y]}{Z} = \Kill{X}{[Y,Z]} \quad \forall \; X, Y, Z \in \g \: .
}
For example,
we could take $\Kill{\cdot}{\cdot}$ to be the Killing form
on $\g$.
For later use, we introduce a non-degenerate Hermitian inner product
$\hip{\cdot}{\cdot}$ on $\g$ defined by
\eqn{hip}{
\hip{X}{Y} := -\Kill{c(X)}{Y} \quad \forall \;  X , Y  \in \g \; ,
}
where
$c:\g \rightarrow \g$ is the conjugation operator
determined by the compact real form $\g_{0}$. From the $\ad$-invariance
of $\Kill{\cdot}{\cdot}$ and the fact that conjugation is
an automorphism of $\g$ it
follows that  $\hip{\cdot}{\cdot}$ satisfies
\alin{hiprel}{
\hip{X}{Y}  &= \overline{\hip{Y}{X}} \; , \\
\hip{c(X)}{c(Y)}  &=  \overline{\hip{X}{Y}} \; , \\
\hip{[X,c(Y)]}{Z}   &= \hip{X}{[Y,Z]} \;
}
for all $X,Y,Z \in \g$. Treating $\g$ as a $\mathbb{R}$-linear space
by restricting scalar multiplication to multiplication by reals,
we can introduce a positive definite inner product
$\rip{\cdot}{\cdot}: \g\, \times \,\g \rightarrow \mathbb{R}$
on $\g$ defined by
\eqn{ripdef}{
\rip{X}{Y} := \text{Re}\hip{X}{Y} \quad \forall \; X , Y \in \g \; .
}
Let $\norm{\cdot}$ denote the norm induced on $\g$ by $\rip{\cdot}{\cdot}$, i.e.
\leqn{norm}{
\norm{X} := \sqrt{\rip{X}{X}} \quad \forall \; X\in \g \; .
}
From the invariance properties satisfied by $\hip{\cdot}{\cdot}$, it
is straightforward to verify that $\rip{\cdot}{\cdot}$ satisfies
\lalign{rip}{
\rip{X}{Y}  &= \rip{Y}{X} \notag \; , \\
\rip{c(X)}{c(Y)}  &=  \rip{X}{Y} \label{rip} \; , \\
\rip{[X,c(Y)]}{Z}   &= \rip{X}{[Y,Z]} \notag \;
}
for all $X,Y,Z \in \g$.

Let $\h$ be a Cartan subalgebra of $\g$ and $R\subset \h^{*}$ the roots
determined by $\h$. Then we have the Cartan decomposition
\eqn{gdecompA}{
\g = \h \bigoplus_{\alpha\in R}  \Cbb \eb_{\alpha}
}
where the nonzero vectors $\eb_{\alpha}$ satisfy
\leqn{eb}{
[H,\eb_{\alpha}] = \alpha(H)\eb_{\alpha} \quad \forall \; H\in \h\: .
}
Note that
\leqn{espace}{
\Cbb \eb_{\alpha} = \{ \: X\in \g \: | \: [H,X] = \alpha(H)X \quad
\forall \; H \in \h \: \} \: .
}
A straightforward consequence of \eqref{espace}, the
Jacobi identity,  and the $\ad$-invariance
of $\Kill{\cdot}{\cdot}$ is that
\leqn{ebe}{
[\eb_{\alpha},\eb_{\beta}] \in \Cbb \eb_{\alpha+\beta}
}
and
\leqn{orthog}{
\Kill{\eb_{\alpha}}{\eb_{\beta}} = 0 \quad \text{ if $\alpha + \beta 
\neq 0$.}
}
Following \cite{k5157}, we define $\tb_{\alpha} \in \h$
as the unique vector in $\h$ that satisfies
\eqn{tal}{
\Kill{\mathbf{t}_{\alpha}}{H}=\alpha(H)\quad \forall\; H\in\h\; .
}
Then
\leqn{Rip}{
\Kill{\alpha}{\beta} := \Kill{\tb_{\alpha}}{\tb_{\beta}} \quad
\forall \; \alpha, \beta \in R
}
defines a positive definite inner product on the
space $\rspan{\{\:\alpha\:|\:\alpha \in R\:\}} \: .$
We will use
$|\cdot|$ to denote the norm of this inner product. 
It will be useful to introduce the  ``dual roots''
$\alpha^{\vee}$ defined by
\eqn{acheck}{
\alpha^{\vee} := \frac{2\alpha\,}{|\alpha|^{2}} \:.
}
We can use the dual roots to define the
angle bracket
\eqn{humpip}{
\humpip{\alpha}{\beta} := \Kill{\alpha}{\beta^{\vee}} \; ,
}
and the vectors
\eqn{h}{
\hb_{\alpha}:= \tb_{\alpha^{\vee}} \; .
}
Choosing a base $\Delta$ for $R$, we then have
\eqn{hdecomp}{
\h = \bigoplus_{\alpha \in \Delta} \Cbb \hb_{\alpha} \:.
}
Also the  Cartan matrix $C$ is defined via
\eqn{C}{
C_{\alpha\beta} := \humpip{\alpha}{\beta} \quad
\forall \; \alpha, \beta \in \Delta \: .
}
A useful relation that is an easy consequence of
the above definitions is 
\leqn{hbe}{
[\hb_{\alpha},\eb_{\beta}] = C_{\beta\alpha}\eb_{\beta}
\quad \forall \; \alpha,\beta \in \Delta \: .
}

Since $\g_{0}$ is a compact real form of $\g$, the vectors
$\{\:\hb_{\alpha},\:\eb_{\alpha}\:|\:\alpha\in R\:\}$ can always be 
chosen to satisfy the 
following relations
\leqn{che}{
c(\hb_{\alpha}) = - \hb_{\alpha} ,\quad
c(\eb_{\alpha}) = - \eb_{-\alpha} \quad \forall \; \alpha \in R\: ,
}
and 
\lalign{cw}{
[\eb_{\alpha}&,\eb_{-\alpha}]  = \hb_{\alpha} \label{cw1}\\
[\eb_{\alpha}&,\eb_{\beta}]  = N_{\alpha,\beta} \eb_{\alpha+\beta}
&& \text{if $\alpha + \beta \in R$} \label{cw2} \\
[\eb_{\alpha}&,\eb_{\beta}]  = 0 && \text{if $\alpha + \beta \neq 0$ 
and $\alpha+\beta \notin R$} \label{cw3}
}
where the constants $N_{\alpha,\beta}$
are real and $N_{\alpha,\beta} = -N_{-\alpha,-\beta}$. We are also
free to normalize the vectors $\eb_{\alpha}$ as follows
\eqn{enorm}{
\Kill{\eb_{\alpha}}{\eb_{\alpha}} = \frac{2}{|\alpha |} \quad
\forall \; \alpha \in R \: .
}
A basis satisfying these conditions will be called
a \emph{Chevalley-Weyl basis}.
The compact real form $\g_{0}$ can then be written as
\eqn{g0decomp}{
\g_{0} = \bigoplus_{\alpha\in\Delta} \Rbb i \hb_{\alpha}\oplus
\bigoplus_{\alpha\in R^{+}} \Rbb\left(\eb_{\alpha}-\eb_{-\alpha}
\right) \oplus 
\bigoplus_{\alpha\in R^{+}} \Rbb i \left(\eb_{\alpha}+\eb_{-\alpha}
\right) \: ,
}
where $R^{+}$ is the set of positive roots. The subspace
\leqn{rcartan}{
\h_{0} = \bigoplus_{\alpha\in\Delta} \Rbb i \hb_{\alpha}
}
is called the real Cartan subalgebra of $\g_{0}$. Notice that
$\h$ is the complexification of $\h_{0}$. As in the complex
case, a real Cartan subalgebra can be defined independently as a maximal
Abelian subalgebra of $\g_{0}$.

From \eqref{rcartan} and the fact that $\Kill{\alpha}{\beta} \in \Rbb$
for all $\alpha, \beta \in R$, it is clear that 
$\alpha(H) \in i\Rbb$ for every $H\in \h_{0}$ and
$\alpha \in R$. This allows us to define a subset $\Wc_{\Rbb}$ of
$\h_{0}$ called the {\em real fundamental open Weyl chamber} by
\eqn{realWeyl}{
\Wc_{\Rbb} := \{ \: H\in \h_{0} \: | \:
-i\alpha(H) > 0 \quad \forall \alpha \in \Delta \: \} \; .
}
We will also need a related subset $\Wc$ of $\h$ called
the {\em (complex) fundamental open Weyl chamber}
which is defined by
\eqn{Weyl}{
\Wc := \{ \: H\in \h \: | \:
\alpha(H) > 0 \quad \forall \alpha \in \Delta \: \} \; .
}
Observe that we have the inclusion $\Wc \subset i\h_{0}$.

If we let $\exp : \g_{0} \rightarrow G$ denote
the exponential map, then the kernel of
$\exp$ is by definition
\eqn{kerexp}{
\ker (\exp) = \{\: X \in g_{0}\: |\: \exp(X) = \id \} \; .
} 
The subset of $\ker(\exp)$ given by
\eqn{integral}{
\Ic := \ker (\exp) \cap \h_{0} \: 
}
is known as the  \emph{integral lattice}.

\sect{sl2}{Three dimensional semisimple Lie subalgebras}

Later on we will see that classifying spherically symmetric
Yang-Mills potentials is related to the problem of
classifying three dimensional semisimple Lie subalgebras of
$\g$ up to
conjugation by inner automorphisms. 
This problem of classifying three dimensional semisimple Lie subalgebras
has been studied extensively by many authors beginning
with Mal'cev \cite{k6436} and Dynkin \cite{k4779}. For a modern presentation
and relations to nilpotent orbits see \cite{k6494}.

It is well known that 
any three dimensional semisimple Lie algebra is isomorphic
to $\Sl{2}$ and is spanned
by three vectors  $\{\:\Oo,\Op,\Om\:\}$
that satisfy the
commutation relationships
\leqn{sl2rel}{
[\Oo,\Omega_{\pm}] = \pm 2\, \Omega_{\pm} \AND
[\Op,\Om] = \Oo \: .
}
The vectors $\{\:\Oo,\Op,\Om\:\}$ are known
collectively as a {\em complex standard triple}.
Instead of working directly with $\Sl{2}$-subalgebras,
we will often find it more convenient to work with $A_{1}$-vectors. An 
{\em $A_{1}$-vector} is a vector $\Oo \in \g$ for which there exists 
two vectors $\Op$, $\Om$
such that the commutation relationships \eqref{sl2rel} are satisfied.
The set of all $A_{1}$-vectors will be denoted by $\Av$. A distinguished
subset of $\Av$ is the set $\rAv$ of { \em real $A_{1}$-vectors}. 
These are the $A_{1}$-vectors $\Oo$ for which
$\Op$, $\Om$ can be chosen so that 
\leqn{r2ctrip1}{
 c(\Oo) = -\Oo \AND c(\Op) = -\Om
}
are also satisfied. Notice that in the real case 
if we define vectors $\Omg{1}$, $\Omg{2}$, and $\Omg{3}$ in $\g_{0}$
via
\leqn{r2ctrip}{
\Omg{0} = 2i\Omg{3} \AND \Omega_{\pm} = \mp \Omg{1} - i \Omg{2} \: ,
}  
then $\Omg{1}$, $\Omg{2}$, and $\Omg{3}$ satisfy
\leqn{so3rel}{
[\Omg{i},\Omg{j}] = \epsilon_{ij}{}^{k}\Omg{k} \; .
}
This shows that $\rspan\{\Omg{1},\Omg{2},\Omg{3}\}$ is isomorphic to
$\So{3}$. Therefore it is appropriate to call
$\{\:\Oo,\Op,\Om\:\}$ a \emph{real standard triple}. 
 
Let
\eqn{Autg}{
\Aut(\g) := \{\:\phi \in \GL(\g) \:|\:\text{ $[\phi(X),\phi(Y)] = \phi([X,Y])$ for all
$X,Y\in \g$}\:\}
}
denote the automorphism group of $\g$. The {\em group of inner automorphisms} $\Int(\g)$
is defined to be the subgoup of $\Aut(\g)$ generated by automorphisms of
the form $\exp(\ad(X))$ where $X$ is any element of $\g$ for which $\ad(X)$ is
nilpotent.  It is a standard
result in Lie algebra theory that $\Int(\g)$ is the identity component of $\Aut(\g)$.
With these conventions we define
\leqn{conjclass1}{
[\Av] :=  \{ \: \text{ $\Int(\g)$ conjugacy classes of $A_{1}$-vectors of
$\g$ }\: \} \: , 
}
and 
\leqn{conjclass2}{
[\Ac_{1}] := \{ \: \text{ $\Int(\g)$ conjugacy classes of $\Sl{2}$-subalgebras
 of $\g$ }\:\}\: .
}
Conjugacy classes of an element $x$ will be denoted by $[x]$. 

It is well known
\cite{k6494} that the map 
\leqn{biject1}{
[\Ac_{1}] \longrightarrow [\Av]\: : \:  [\cspan\{{\Oo,\Op,\Om}\}] \longmapsto [\Oo]
}
is a bijection.  In \cite{k4779} Dynkin proved that for a fixed Cartan subalgebra $\h$ and
base $\Delta = \{\: \alpha_{1},\alpha_{2},\ldots,\alpha_{\ell}\:\}$ 
there exists a unique $A_{1}$-vector $\Oo$ in a conjugacy class
$[\Oo']$ such that
\eqn{Dynkin}{
\text{$\alpha(\Oo) =  0$, $1$, or $2$ for all $\alpha \in \Delta$.}
}
He then defined the {\em characteristic} $\chi([\Oo'])$ of the conjugacy
class $[\Oo']$ by
\eqn{Aasymp}{
\mathbf{\chi} =
 \chi([\Oo']) := \bigl(\alpha_{1}(\Oo),\ldots,\alpha_{\ell}(\Oo)\bigr)\: .
}
The importance of the characteristic is that it is a complete
invariant, i.e.  $[\Oo''] = [\Oo']$ if and only if $\chi([\Oo''])
= \chi([\Oo'])$. Consequently,
\leqn{biject2}{
\text{ $\Av\cap \overline{\mathcal{W}}$ is in one-to-one correspondence with
$[\Av]$.} 
} 
For our purposes, we are more interested in the set 
$\rAv \cap \overline{\mathcal{W}}$. We then have the
useful result:
\begin{lem} \label{rA1lem} \mnote{[rA1lem]}
\eqn{rA1lem1}{
\rAv \cap \overline{\mathcal{W}} = \Av\cap \overline{\mathcal{W}}
}
\end{lem}
\begin{proof}
Since $\rAv \cap \overline{\mathcal{W}} \subset \Av\cap \overline{\mathcal{W}}$
we only need to verify the reverse inclusion. So assume that $\Oo \in \Av\cap 
\overline{\mathcal{W}}$. Then there exists $\Op,\Om \in \g$ such that
$\Sc := \cspan \{\Oo,\Op,\Om\} \cong \Sl{2}$. Let $\Omg{1}$, $\Omg{2}$, and $\Omg{3}$
be as in \eqref{r2ctrip}. Then $\Sc_{0} := \rspan\{\Omg{1},\Omg{2},\Omg{3}\}$
is a compact real form for $\Sc$. Now $\Sc_{0}$ sits inside some maximal compact
real form $\g_{0}' \subset \g$. But $\g_{0}$ is also a maximal compact real form and
as all maximal compact real forms  are conjugate
by inner autmorphism (see theorem 2.1 p. 256 in \cite{k1944}) 
there exist a $\sigma \in \Aut(\g)$ such that
$\sigma(\Sc_{0}) \in \g_{0}$. In particular $\sigma( \Omg{3}) \in \g_{0}$.
We also have that $\Omg{3} \in \g_{0}$ and hence it follows 
by proposition 8.3.1 of \cite{k6468} that 
$\Ad_{g}(\Omg{3}) = \sigma (\Omg{3})$ for some $\g\in G$. Thus 
$\cspan\{\Oo, \Ad_{g^{-1}}\circ \sigma (\Op), \Ad_{g^{-1}}\circ \sigma(\Om)\}$ is
a real standard triple.
\end{proof}
It is worthwhile to note that not every combination
of $0$, $1$, and $2$ defines the characteristic of some conjugacy class
$[\Oo]$. In fact, the total number of conjugacy classes is far
less than the potential $3^{\ell}$. For example, 
the number of characteristics for $A_{\ell-1}$ is equal to to
the number of partitions of $\ell$ and this is asymptotically equivalent
to 
\eqn{partitions}{
\frac{1}{4\sqrt{3}\ell} e^{\pi\sqrt{\frac{2\ell}{3}}} \; ,
}  
which is much smaller than $3^{\ell-1}$. 

It is not difficult
to show that for every Lie algebra $\g$ there is always
a characteristic of the form
\eqn{chidistg}{
\chi = (2,2,\ldots,2) \: .
}
In other words there always exists an  $A_{1}$-vector $\Oo$ such
that $\alpha(\Oo) = 2$ for all $\alpha \in \Delta$. These
distinguished elements will be called {\em principal $A_{1}$-vectors}.

The Dynkin diagram of a Lie algebra $\g$ 
labeled with the characteristic numbers $\alpha_{k}(\Oo)$
above the nodes is called a {\em weighted Dynkin diagram}.
All the possible weighted Dynkin diagrams of the exceptional Lie algebras
$G_{2}$, $F_{4}$, $E_{6}$ and $E_{8}$ were determined by Dynkin
in \cite{k4779}. A listing of these diagrams can be found
in \cite{k6494} section 8.4 .

For the classical Lie algebras the weighted Dynkin diagrams are not
optimal for classifying the conjugacy classes of $\Avc$. Instead,
a different method based on the ``partitions of n'' is
used. To describe
this method, we first consider $\Sl{n} = A_{n-1}$ for
which the classification problem can be solved by elementary
methods.  A {\em partition } of $n$ is an $k$-tuple
$\db := (d_{1},d_{2},\ldots,d_{k})$ such that
\leqn{partition}{
d_{1} \geq d_{2} \geq \ldots \geq d_{k} > 0\:
\AND  n = \sum_{j=1}^{k} d_{j} \: .
}
If a number $s$ is repeated $q$ times in a partition we will
denote this by $s^{q}$ and $q$ will be called the
{\em multiplicity} of $s$. For example, the  partition 
$(9,9,9,6,4,4,2,1,1,1)$ will also be written as
$(9^{3},6,4^{2},2,1^{3})$.
The set of all partitions of $n$ will be denoted by $\Pc(n)$.
Using $\Sl{2}$ representation theory,  it not hard to show 
that the there exists a bijection from $\Avc$ to $\Pc(n)$. 
Moreover, for each partition $(d_{1},\ldots,d_{k})$ a canonical
representative
$\Oo^{(d_{1},\ldots,d_{k})}$
of the conjugacy class can be constructed as
follows. For each $s \in \mathbb{N}$ let 
\leqn{Ablock1}{
\Oo^{s} = \begin{pmatrix}
s &  0  &  0  & \cdots & 0 & 0 \\
0 & s-2 &  0  & \cdots & 0 & 0 \\
0 &  0  & s-4 &        & 0 & 0 \\
\vdots & \vdots &  & \ddots & & \vdots \\
0 & 0 & 0 &  & -s+2 & 0 \\
0 & 0 & 0 & \cdots & 0 & -s
\end{pmatrix} 
 \; .}
Then
\leqn{Ablock2}{
\Oo^{(d_{1},\ldots,d_{k})} = \bigoplus_{j=1}^{k} \Oo^{d_{j}-1}\: 
}
is the canonical representative. There also exists
simple formulas for $\Omega_{\pm}$.
For each $s \in \mathbb{N}$ let
\leqn{Ablock3}{
\Op^{s} = \begin{pmatrix}
0       &\sqrt{1(s)} &        0          &  0            & \cdots & 0      \\
0       & 0          &     \sqrt{2(s-1)} &  0            & \cdots & 0      \\ 
0       &   0        &        0          & \sqrt{3(s-2)} &        &       \\
0       &  0         &        0          &    0          &        &       \\
\vdots  & \vdots     &      \vdots       &               & \ddots &  \\
0       &0           &        0          &       0       &  & \sqrt{(s)1}      \\
0       &0           &        0          & 0             & \cdots &  0  
\end{pmatrix}
 \; }
and
\leqn{Ablock4}{
\Om^{s} = (\Op^{s})^{\text{t}}
} 
where ${}^{\text{t}}$ denotes the transpose of a matrix.
Then
\leqn{Ablock5}{
\Omega^{(d_{1},\ldots,d_{k})}_{\pm} = \bigoplus_{j=1}^{k} 
\Omega^{d_{j}-1}_{\pm}\: .
}
From these formulas it is easy to verify that
\eqn{slntrip}{
\cspan\{\Oo^{(d_{1},\ldots,d_{k})},
\Op^{(d_{1},\ldots,d_{k})},\Om^{(d_{1},\ldots,d_{k})}\} \cong \Sl{2} \, .
}
Similar results can be obtained for the other classical
Lie algebras, with the conjugacy classes of $\Avc$ being
parametrized by a subset of $\Pc(n)$ for appropriate $n$.
A canonical representative of the
conjugacy class can also be constructed although the formulas
are more complicated. All of this can be found
in chapter 5 of \cite{k6494}, we only state the results.

\begin{thm}\label{sl2thm3} \mnote{sl2thm3]}
If $\g = \Sl{n}$ then there  exists a bijection 
between $[\Ac_{1}^{\text{\emph{v}}}]$ and $\Pc(n)$.
\end{thm}

\begin{thm}\label{sl2thm4} \mnote{sl2thm4]}
If $\g = \mathfrak{so}_{2n+1}\Cbb$ then there  exists a bijection 
between $[\Ac_{1}^{\text{\emph{v}}}]$ and the set of partitions of $2n + 1$
in which the even parts occur with even multiplicity.
\end{thm}
Example: $\mathfrak{so}_{7}\Cbb$ contains 
six conjugacy classes parametrized by the
partitions  $(7)$, $(5,1^{2})$, $(3,1^{4})$,
$(3,2^{2})$, $(3^{2},1)$, and $(2^{2},1^{3})$.

\begin{thm}\label{sl2thm5} \mnote{sl2thm5]}
If $\g = \mathfrak{sp}_{2n}\Cbb$ then there  exists a bijection
between $[\Ac_{1}^{\text{\emph{v}}}]$ and the set of partitions of $2n$
in which the odd parts occur with even multiplicity.
\end{thm}
Example: $\mathfrak{sp}_{6}\Cbb$ contains
seven conjugacy classes parametrized by the
partitions  $(6)$, $(4,2)$, $(4,1^{2})$,
$(3^{2})$, $(2^{3})$, $(2^{2},1^{2})$, and $(2,1^{4})$.

\begin{thm}\label{sl2thmr6} \mnote{sl2thm7]}
If $\g = \mathfrak{so}_{2n}\Cbb$ then there  exists a bijection
between $[\Ac_{1}^{\text{\emph{v}}}]$ and the set of partitions of $2n$
in which the even parts occurs with even multiplicity 
except that the ``very even'' partitions 
$(d_{1},\ldots,d_{k})$( those
with only even parts, each having even multiplicity)
correspond to conjugacy classes labeled
$(d_{1},\ldots,d_{k})_{I}$ and
$(d_{1},\ldots,d_{k})_{II}$. 
\end{thm}
Example: $\mathfrak{so}_{8}\Cbb$ contains
eleven conjugacy classes parametrized by the partitions
$(7,1)$, $(5,3)$, $(4^{2})_{I}$, $(4^{2})_{II}$ ,
$(5,1^{3})$, $(3^{2},1^{2})$, $(3,2^{2},1)$,
$(2^{4})_{I}$, $(2^{4})_{II}$, $(3,1^{5})$,
and $(2^{2},1^{4})$.

\sect{eymeqns}{Static spherically symmetric field equations}

Let $P$ be a principal bundle with a compact semi-simple structure group $G$
over a static spherically symmetric space-time manifold. We 
consider only actions of the group $SU(2)$ by principal bundle automorphisms
on $P$ that project onto the action of $SO(3)$ on space-time which defines the
spherical symmetry. This ignores some interesting effects due to the
fact that $SO(3)$ is not simply connected. For an analysis of $SO(3)$
actions on $SU(n)$ bundles see \cite{k6217}.

Equivalence classes of these spherically symmetric $G$-bundles are in
one-to-one correspondence to conjugacy classes of homomorphisms of the
isotropy subgroup, $U(1)$ in this case, into $G$. 
The $U(1)$ conjugacy classes are classified by the following
proposition.
\begin{prop}{\emph{[proposition 1,\cite{k5281}]}} 
\label{gpotthm1} \mnote{[gpotthm1]}
The set of conjugacy classes of homomorphisms $\lambda : U(1)
\rightarrow G$ is in one-to-one correspondence with
the set $\Ic \cap \overline{\Wc}_{\Rbb}$. The conjugacy class of $\lambda$
is characterized by $\lambda'(\eb) \in \Ic \cap \overline{\Wc}_{\Rbb}$
where $\lambda': \mathfrak{u}(1) \rightarrow \g_{0}$
is the induced Lie algebra homomorphism
and $\eb = 2\pi i$ is the standard basis vector in
the integral lattice of $\mathfrak{u}(1)$.
\end{prop}
So, once a principal $G$-bundle with an $SU(2)$-action is fixed,
it determines a element
$\Ic \cap \overline{\Wc}_{\Rbb}$ which we will denote as
$-4\pi \La_{3}$. 

Wang's theorem \cite{k4872,k0929} on connections that are invariant
under actions transitive on the base manifold has been adapted to spherically
symmetric space-time manifolds by Brodbeck and Straumann \cite{k5281}. They
show that in a Schwarzschild type coordinate system $(t,r,\theta,\phi)$ and
the metric
\leqn{statmetric}{g = -N(r) S(r)^{2} dt^{2}+
  N(r)^{-1}dr^{2}+r^{2}(d\theta^{2}+\sin^{2}\theta d\phi^{2}) \; .
}
a gauge can always be chosen such that the $\g_{0}$-valued Yang-Mills connection
form is locally given by
\leqn{statconnx}{
A=\tilde{A}+\hat{A}
}
where
\leqn{statAtilde}{
\tilde{A} = N(r)S(r)\Ac(r) dt + \Bc(r) dr
}
is a $\g_{0}^{\La_{3}}$-valued 1-form, and
\leqn{Ahatstat}{
  \hat{A} = \La_{1}(r) d\theta +
            ( \La_{2}(r)\sin\theta + \La_{3}\cos\theta)d\phi
}
where $\La_{1}$ and $\La_{2}$ are $\g_{0}$-valued maps that satisfy
\leqn{wang}{
[\La_{2},\La_{3}]=\La_{1} \AND
[\La_{3},\La_{1}]=\La_{2}.
}
By the results of \cite{k6643} \S 3.3 we are free to use
the temporal gauge and therefore set
\leqn{stattemp}{
 \Bc = 0 \; .
}
We will make one more assumption on the form of the gauge potential,
namely that $\Ac = 0$. In analogy with the electromagnetic theory,
we call $\tilde{A}$ and $\hat{A}$ the electric and magnetic parts
of the gauge potential, respectively. Thus \eqref{stattemp} and the assumption
$\Ac =0$ means that the gauge potential is purely magnetic. 
For solutions that are bounded at the origin, it can
be shown by analyzing the initial value problem at
$r=0$ using the techniques \cite{eymg,eymgg} that $\Ac = 0$ is
a consequence of the EYM equations.
Therefore no generality is lost by setting $\Ac = 0$ when
looking for solutions that are bounded at $r=0$. However,
for black hole solutions it is known that there exists solutions
with $\Ac$ not identically zero \cite{k5187,k6174}. Therefore $\Ac = 0$ 
is a restriction in this case.

With the above assumption, the gauge potential takes the form
\leqn{connx1}{
A =  \La_{1}(r) d\theta +
            ( \La_{2}(r)\sin\theta + \La_{3}\cos\theta)d\phi \: .
}
Using \eqref{statmetric} and \eqref{connx1}, the EYM equations
reduce to
\lgath{feq}{
m' = (NG + r^{-2}P),                         \label{feq1}\\
S^{-1}S' = 2 r^{-1}G,                        \label{feq2}\\
r^{2}N \Lp'' + 2(m-r^{-1}P)\Lp' + \Fc = 0,   \label{feq3}\\
[\Lp',\Lm] + [\Lm',\Lp] = 0                  \label{feq4}
}
where ${}' := d/dr$ and
\lgath{vardefs}{
\Lpm := \mp \Lambda_{1} - i \Lambda_{2}, \quad  \Lo := 2i\La_{3} , 
\label{vardefs2} \\ 
N   =: 1 - \frac{2m}{r},  \quad
G   := \half \hip{\Lp'}{\Lp'} \quad P := \half \hip{\Fh}{\Fh}, \label{vardefs3} \\
\Fh := \ihalf(\Lo-[\Lp,\Lm]),      \label{vardefs4} \\
\Fc := -i[\Fh,\Lp].                \label{vardefs5}
}
Using the norm \eqref{norm}, $G$ and $P$ can 
can be written as
\leqn{vardefsa}{
G   = \half \|\Lp'\|^{2} \AND P = \half \|\Fh\|^{2}
}
and  obviously $G\ge0$ and $P\ge0$. 
A useful variant of \eqref{feq1} is
\leqn{feq1v}{
N' = \frac{1}{r}\left(1-N-2NG-\frac{2}{r^{2}}P\right)\: .
}
Observe that equation \eqref{wang} becomes
\leqn{nwang}{
[\Lo,\Lpm] = \pm 2 \Lpm\:,
}
and $\Lo$ satisfies
\leqn{cLo}{
c(\Lo) = -\Lo \: .
}
Defining 
\leqn{Slam}{
S_{\lambda} := \{\: \alpha \in R \: | \: \alpha(\Lo) = 2 \: \}
}
it follows easily from \eqref{nwang} that
\leqn{Lpim}{
\Lp(r) \in \bigoplus_{\alpha \in S_{\lambda}} \Cbb \eb_{\alpha} \quad \forall \; r\: .
}
Observe that if we define
\leqn{Vn}{
V_{n} := \{ \, X\in \g \, | \, [\Lo,X] = n X \, \} \, ,
}
then
\leqn{V2Slam}{
V_{2} := \bigoplus_{\alpha \in S_{\lambda}} \Cbb \eb_{\alpha}
}

It is straightforward to verify that the
stress-energy tensor $T^{a}_{b}$ in the Schwarzschild coordinates
is given by 
\eqn{diagstren}{
T^{a}_{b} = \diag\left(-
\frac{NG}{r^{2}}-\frac{P}{r^{4}},
\frac{NG}{r^{2}}-\frac{P}{r^{4}},
\frac{P}{r^{4}}, \frac{P}{r^{4}}\right)
}
So the energy
density, radial and tangential pressure are given by
\leqn{ener}{
 4\pi e = r^{-2}(NG+r^{-2}P), \quad
 4\pi p_{r} = r^{-2}(NG-r^{-2}P), \AND
 4\pi p_{\theta} = r^{-4}P,
}
respectively.

The magnetic and electric charges can be defined via
\eqn{tYMcharge}{
Q_{E} :=  
\ensuremath{\textstyle\frac{1}{4\pi}} \int_{\Sc^{2}_{\infty}} \norm{*F_{ab}\epsilon
^{ab}}\epsilon \AND
Q_{M} := \ensuremath{\textstyle\frac{1}{4\pi}}
\int_{\Sc^{2}_{\infty}} \norm{F_{ab}\epsilon^{ab}}\epsilon \: .
}
where $\Sc^{2}_{\infty}$ is a sphere at $\infty$ and $\epsilon$ is
the induced volume form on $\Sc^{2}_{\infty}$. 
Using \eqref{connx1}, a short calculation shows that total magnetic
and electric charges are given by
\leqn{tYMcharge1}{
Q_{M} = \lim_{r\rightarrow \infty} \|\Fh(r)\|
\AND Q_{E} = 0 \: .
}

Following \cite{k5428}, we find it useful to introduce
a new independent variable $\tau$ via
\leqn{tau}{
\frac{dr}{d\tau} = r\sqrt{N}
}
and dependent variables
\leqn{BFMvars}{
\mu := \sqrt{N},\quad \Up := \sqrt{N}\Lp', \AND
\kappa := \frac{1}{2\mu} ( 1+\mu^{2}+2\mu^{2} G - 2r^{-2}P) \; .
}
In these variables,  equations \eqref{feq1}-\eqref{feq4} become
\lgath{feqBFM}{
\dot{r} = r\mu \: ,                          \label{feqBFM1}\\
\dot{\Lambda}_{+} = r \Up\: ,                        \label{feqBFM2}\\
\dot{\mu} = (\kappa-\mu)\mu - 2\mu^{2} G\: ,   \label{feqBFM3}\\
(S\mu)\dot{} = S(\kappa-\mu)\mu\:,                  \label{feqBFM4}\\
\dot{\kappa} = -\kappa^{2}+1+2\mu^{2}G\: ,   \label{feqBFM5} \\
\dot{U}_{+} = -(\kappa-\mu)\Up - \frac{1}{r}\Fc \: , \label{feqBFM6}
}
where $\dot{(\cdot)} = \frac{d(\cdot)}{d\tau}$. One advantage of
this system of equations over \eqref{feq1}-\eqref{feq4} is
that it is no longer singular at $\mu = N^{2} = 0$. However,
this will not be important to us here. Instead, we shall
exploit in section \ref{aym} the fact that the system \eqref{feqBFM1}-\eqref{feqBFM6}
is asymptotically autonomous to determine the behaviour of bounded solutions as
$r \rightarrow \infty$. 

\sect{restLo}{Restrictions on $\Lo$}

Spherical symmetry implies that $\Lo$ must lie in the set 
$\frac{1}{2\pi i}\Ic\cap \overline{\Wc}_{\Rbb}$. 
We shall now see how boundary conditions restrict
$\Lo$ to lie in an even smaller set. 
For the geometry to be regular at the origin, it is
necessary that
\leqn{reggeom}{
\lim_{r\rightarrow 0} N(r) = 1 \: .
}
For a global solution that is defined for 
all $r \in [0,\infty)$, the physical
boundary conditions at the origin $r=0$ are that the
energy density and the radial and tangential
pressures are finite there. From \eqref{ener} and
\eqref{reggeom}, is clear that these boundary conditions
 imply 
\leqn{originbcs}{
G(r) = \text{O}(r^{2}) \AND P(r) = \text{O}(r^{4}) \quad
\text{as $r\rightarrow 0$} \: .
} 
An immediate consequence is
\eqn{sl2orig}{
[\Lp(0),\Lm(0)] = \Lo \: .
}
This result combined with \eqref{nwang} and \eqref{cLo} shows
that $\{ \Lo,\Lp(0),\Lm(0)\}$
form a real standard triple. Thus $\Lo \in \rAv\cap 
\left(\frac{1}{2\pi i}\Ic\cap \overline{\Wc}_{\Rbb} 
\right)$.
\begin{lem}  \label{rLolem1} \mnote{[rLolem1]}
\eqn{rLolem1.1}{
\Ac_{1}^{\text{\emph{v}}, \Rbb} \cap \overline{\Wc} = 
\Ac_{1}^{\text{\emph{v}}, \Rbb} \cap \left( 
\frac{1}{2\pi i}\Ic\cap \overline{\Wc}_{\Rbb} \right)
}
\end{lem}
\begin{proof}
Suppose $\Oo \in \rAv \cap \Wc$.
Then  $2\pi i\Oo \in \Wc_{\Rbb}$. Moreover, $\exp(2\pi i \Oo) = 1 \in G$ 
as $\Oo$
is the neutral element of a $\Sl{2}$ subalgebra and 
$2\pi i \Lo \in \h_{0}$. Therefore $\Oo \in \frac{1}{2\pi i}\Ic\cap \overline{\Wc}_{\Rbb}$
and hence
\eqn{rLolem1.2}{
\rAv \cap \overline{\Wc} \subset \rAv \cap \left( 
\frac{1}{2\pi i}\Ic\cap \overline{\Wc}_{\Rbb} \right) \: .
}
The reverse inclusion is simple to establish and will
be left to the reader.
\end{proof}
This lemma shows that with the above boundary conditions
at the origin $r=0$, we can assume that 
\leqn{reduced}{
\Lo \in \rAv \cap \overline{\Wc} \: .
}

As we remarked before, the assumption
that the spacetime is asymptotically flat implies that
\leqn{flat}{
\lim_{r\rightarrow \infty } N(r) = \lim_{r\rightarrow \infty} S(r) = 1 \: .
}
A common boundary condition that is adopted at $r = \infty$ is
that the total magnetic charge vanishes. 
The  vanishing of the total magnetic charge is equivalent to 
\leqn{QMvanish}{
\lim_{r\rightarrow \infty } \Fh = 0 \: 
}
by \eqref{tYMcharge1}.
Assuming the limit $\lim_{r\rightarrow \infty} \Lp(r)$ exists,
\ref{QMvanish} implies
\eqn{sl2infty}{
[\Lp(\infty),\Lm(\infty)] = \Lo \: 
}
where $\Lp(\infty) = \lim_{r\rightarrow \infty} \Lp(r)$.
The same argument as above
shows that $\Lo \in \rAv \cap \overline{\Wc}$. Therefore,
if the magnetic charge vanishes,
then we can assume that $\Lo \in \rAv \cap \overline{\Wc}$.

The condition  \eqref{QMvanish} does
not seem to be a necessary one as purely magnetic black hole
solutions have been constructed numerically with nonzero magnetic
charge \cite{k6174}. However, for globally regular solutions
defined on $[0,\infty)$ it is unknown if this condition
is necessary.  Indeed, as we shall see in section \ref{chgb} for certain 
choices of $\Lo \in \rAv \cap \overline{\Wc}$, it is not clear that
the limit $\lim_{r\rightarrow \infty} \Lp(r)$ actually exists, and even
if it does exists we have not been able to prove that \eqref{QMvanish} is 
automatically satisfied. With this said, we will for the remainder of
the article assume that \ref{reduced} holds.

\sect{regA1}{Regular $A_{1}$-vectors and $\Pi$-systems}

If $\Lo \in \rAv \cap \Wc$ then we call $\Lo$ a {\em regular 
$A_{1}$-vector} and the action of $SU(2)$ determined by
$\Lo$ will be called a \emph{regular action}. 
Previous to our work in \cite{eymgg}, 
all the results in the literature
concerning the EYM equations have be derived under the
assumption that $\Lo$ is regular. There are two main
reasons for this assumption. First, equation \eqref{feq4} can be solved 
exactly and secondly the remaining equations \eqref{feq1}-\eqref{feq3}
can be expanded out in a Chevalley-Weyl  basis $\{\: \hb_{\alpha} \: | \:
\alpha \in \Delta \:\}\cup \{\: \eb_{\alpha} \: | \: \alpha \in
R \: \}$ without having to explicitly compute any of 
the brackets $[\eb_{\alpha},\eb_{\beta}]$. 
As we shall see below, this simplification 
can be traced back to the fact
that $S_{\lambda}$ is a 
{\em $\Pi$-system} whenever $\Lo$ is regular.  So in fact the simplification is not
dependent on $\Lo$ being regular but only on 
$S_{\lambda}$ being a $\Pi$-system. We recall \cite{k4779}
that a subset $ \Sigma \subset R$ is called a {\em $\Pi$-system}
if and only if 
\begin{itemize}
\item[(i)] if $\alpha,\beta \in \Sigma$ then $\alpha-\beta \notin R$
\item[(ii)] $\Sigma$ is linearly independent
\end{itemize}
For a proof of the fact that $\Lo$ regular implies that
$S_{\lambda}$ is a $\Pi$-system see \cite{k4295}.

If we assume that $S_{\lambda}$ is a $\Pi$-system, then
$\{\: \hb_{\alpha}\, ,\, \eb_{\alpha}\, ,\, \eb_{-\alpha} \: | \: \alpha \in S_{\lambda}\:\}$ generates a 
semisimple Lie subalgebra of
$\g$ denoted $\g_{\lambda}$ for which $S_{\lambda}$ is
a base \cite{k4779}. By the definition of $S_{\lambda}$, it is clear that
$\Lo$ is a principal $A_{1}$-vector in $\g_{\lambda}$. Also
from $\eqref{Lpim}$ and the definition of $\g_{\lambda}$, we
see that $\Lp(r) \in \g_{\lambda}$ for all $r$. The
above discussion shows that if $\Lo \in \rAv \cap \overline{\Wc}$
is chosen so
that $S_{\lambda}$ is a $\Pi$-system, then the
field equations \eqref{feq1}-\eqref{feq4}
can be reduced to a subalgebra of $\g$ for which $\Lo$ is a 
principal. Therefore, when $S_{\lambda}$ is 
a $\Pi$-system we can, without loss of generality, assume
that $\Lo$ is a principal $A_{1}$-vector in $\g$.

So assume now that $\Lo  \in \rAv \cap \overline{\Wc}$ is
principal. 
We have the expansion
\leqn{Lpexp}{
\Lp(r) = \sum_{\alpha\in S_{\lambda}} w_{\alpha}(r) \eb_{\alpha} \; , 
}
by \eqref{Lpim}
where the $w_{\alpha}(r)$ are complex valued functions and
$\Delta = S_{\lambda}$ .
From
\eqref{che} and \eqref{vardefs2} it follows that
\leqn{Lmexp}{
\Lm(r) = \sum_{\alpha\in S_{\lambda}} \bar{w}_{\alpha}(r) \eb_{-\alpha} \; .
}
Substituting \eqref{Lpexp} and \eqref{Lmexp} into \eqref{feq4}
and using \eqref{cw1}-\eqref{cw3} yields
\leqn{csrteqn}{
w_{\alpha}\bar{w}_{\alpha}'= w_{\alpha}'\bar{w}_{\alpha} \quad
\forall \; \alpha \in  S_{\lambda}
}
since $\alpha$, $\beta \in 
S_{\lambda}$ with $\alpha \neq \beta$ implies that $\alpha-\beta \notin R$. 
Solving equation \eqref{csrteqn} shows that $w_{\alpha}$ must have
constant phase. We are free to choose these phases, which amounts 
to a choice of gauge, and so we demand that the 
phases are all zero. Hence the $w_{\alpha}(r)$ are all real valued functions.
We can substitute \eqref{Lpexp} and \eqref{Lmexp} into 
\eqref{feq1} and \eqref{feq3} to get
\lgath{neqs}{ m' = (NG + r^{-2}P), \label{neqs1}\\
  r^{2} N w_{\alpha}'' + 2(m-r^{-1}P)w_{\alpha}' + \half
  \DS\sum_{\beta \in S_{\lambda} } w_{\alpha} C_{\alpha \beta}
(\lambda_{\beta}-w_{\beta}^{2}) = 0\label{neqs2}
}
where $(C_{\alpha \beta }):=(\langle \alpha, \beta \rangle)$ 
is the Cartan matrix of the reduced structure
group, and
\lalign{nPG}{ P &= {\textstyle\frac{1}{8}}
  \sum_{\alpha, \beta \in S_{\lambda}} (\lambda_{\alpha}-w_{\alpha}^{2}) 
h_{\alpha \beta} (\lambda_{\beta}-w_{\beta}^{2}) \, ,
  \label{nPG1}\\
  G &= \sum_{\alpha \in S_{\lambda} }\frac{{w_{\alpha}'}^{2}} { |\alpha|^{2} } \, ,\label{nPG2}\\
  h_{\alpha \beta} &= \frac{2 C_{\alpha\beta}}{|\alpha|^{2}} \, , \label{nPG3} \\
\intertext{and}
  \lambda_{\alpha} &= 2 \sum_{\beta\in S_{\lambda}}
(C^{-1})_{\alpha \beta} \, .\label{nPG4}
}
As before, in deriving the above expression we have used
$\alpha$, $\beta \in S_{\lambda}$ 
with $\alpha \neq \beta$,
implies $\alpha-\beta \notin R$. 

On the other hand, if $S_{\lambda}$ is not a $\Pi$-system       
then equation  \eqref{feq4} can no longer be solved exactly. 
This is due to the fact that for $\alpha$,$\beta \in R$ with
$\alpha \neq \beta$ it is no longer necessary that $\alpha - \beta
\notin R$. This implies in particular that the bracket 
$[\eb_{\alpha},\eb_{\beta}]$ may no longer be zero. The inability
to solve \eqref{feq4} implies that the system of equations 
\eqref{feq1}-\eqref{feq4} can
no longer be written in the standard form $ y'(r) = f(y(r),r)$
which provides a serious complication. Also the non-vanishing
of the brackets $[\eb_{\alpha},\eb_{\beta}]$ greatly
increases the complexity of the equations.

In view of the above discussion, it would be desirable
to classify all those $\Lo \in \rAv \cap \overline{\Wc}$ for
which $S_{\lambda}$ is a $\Pi$-system.
Suppose $\Lo \in \rAv
\cap \overline{\Wc}$ is such that $S_{\lambda}$ is a $\Pi$-system.
Let  $\g = \bigoplus_{j} \g^{j}$ denote the decomposition of
$\g$ into simple ideals and $R^{j} \subset R$ denote the roots
of $\g^{j}$. This determines a decomposition
of $S_{\lambda} = \dot{\cup} S^{j}_{\lambda}$ into a disjoint
union of sets $ S^{j}_{\lambda} \subset R^{j}$ such that $ S^{j}_{\lambda}$
is a $\Pi$-system in $\g^{j}$. Moreover, if we let
$\Lo = \sum_{i} \Lo^{j}$ denote the corresponding decomposition
of $\Lo$ then it is not difficult to show that
$S^{j}_{\lambda} = \{\: \alpha \in R^{j} \: | \: \alpha(\Lo^{j})=2 \: \}$
and $\Lo^{j} \in  \rAv(\g^{j}) \cap \overline{\Wc}(\g^{j})$. 
This proves that if we can parametrize the set
$\{\: \Lo \in \rAv
\cap \overline{\Wc} \: | \: \text{ $S_{\lambda}$ is
a $\Pi$-system}\}$ for simple Lie algebras $\g$ then
we can parametrize it for all semisimple Lie algebras. The
next theorem provides such a parametrization for the
simple Lie algebras.

\begin{thm}\label{regA1thm1} \mnote{regA1thm1]}
${\;}$\\ ${\; }$ \\ ${\;}$
$\g= \Sl{n}$
\\
$S_{\lambda}$ is a $\Pi$-system if and only if the partition
$\db$ that determines $\Lo$ satisfies one of
the following
\begin{enumerate}
\item[(i)]
$\quad \db = (2v,2u+1)$ with $2v > 2u+1 \geq 3$, 
\item[(ii)]
$\quad \db = (2u+1,2v)$ with $2u+1 > 2v \geq 2$,
\item[(iii)]
$\quad \db = (2v,1,1,\ldots,1)$ with $v \geq 1$.
\end{enumerate}
$\g= \mathfrak{so}_{2n+1}\Cbb$\\
\\
$S_{\lambda}$ is a $\Pi$-system if and only if the partition
$\db$ that determines $\Lo$ satisfies one of
the following
\begin{enumerate}
\item[(i)]
$\quad \db = (2u+1)$ with $u \geq 1$,
\item[(ii)]
$\quad \db = (2u+1,2v,2v)$ with $u \geq 1$, and  $2u+1>2v$,
\item[(iii)]
$\quad \db = (2v,2v,2u+1)$ with $u \geq 1$ and $2v>2u +1$, 
\item[(iv)] $\quad \db = (2v,2v,1,1,\ldots,1)$ with
$v\geq 1$.
\end{enumerate} 
$\g= \mathfrak{sp}_{2n}\Cbb$\\
\\
$S_{\lambda}$ is a $\Pi$-system if and only if the partition
$\db$ that determines $\Lo$ is of the form
$\db = (2v,1,1,\ldots,1)$ where $v\geq 1$. \\
\\
$\g= \mathfrak{so}_{2n}\Cbb$\\
\\
$S_{\lambda}$ is a $\Pi$-system if and only if the partition
$\db$ that determines $\Lo$ satisfies one of
the following
\begin{enumerate}
\item[(i)]
$\quad \db = (u,u)$ with $u \geq 1$,
\item[(ii)]
$\quad \db = (2u+1,1)$ with $u \geq 1$, 
\item[(iii)]
$\quad \db = (2u+1,2v,2v,1)$ with $2u+1 > 2v \geq 2$,
\item[(iv)] $\quad \db = (2v,2v,2u+1,1)$ with $2v > 2u+1 \geq 1$,
\item[(v)] $\quad \db = (2v,2v,2u+1,2u+1)$ with $2v > 2u+1 \geq 3$,
\item[(vi)] $\quad \db = (2u+1,2u+1,2v,2v)$ with $2u+1 > 2v \geq 2$.
\end{enumerate}
\bigskip

\noindent $S_{\lambda}$ is a $\Pi$-system if and only if the characteristic
$\chi$ that determines $\Lo$ is equal to
\begin{itemize}
\item[(a)] $(2^{6})$, $(2,1^{2},0,1,2)$ or $(0,1,0^{4})$ for $\g = E_{6}$,
\item[(b)] $(2^{7})$ or $(2,1,2,1,0,1,2)$ for $\g = E_{7}$,
\item[(c)] $(2^{8})$, $(2,1,2^{2},1,0,1,2)$ or $(0^{7},1)$ for $\g = E_{8}$,
\item[(d)] $(2^{4})$, $(1,0,1,2)$ or $(1,0^{3})$ for $\g = F_{4}$,
\item[(e)] $(2^{2})$, $(1,0)$ or $(0,1)$ for $\g = G_{2}$.
\end{itemize}
\end{thm}
\begin{proof}

From section \ref{sl2}, we know that the sets $\Av \cap  \overline{\Wc}$
can be completely parametrized for the simple Lie algebras. 
Since $\Av \cap  \overline{\Wc}
= \rAv \cap \overline{\Wc}$ by lemma \ref{rA1lem}, 
we can use this parametrization to determine all the 
$\Lo \in \Av \cap  \overline{\Wc}$ such
that $S_{\lambda}$ is a $\Pi$-system. For the exceptional algebras
this can be done by straightforward calculations using 
the tables in chapter 8 of \cite{k6494}.
For the classical algebras, 
we will only prove the theorem for simplest case $\g = \Sl{n}$. The other
algebras $\mathfrak{so}_{2n+1}\Cbb$,  $\mathfrak{so}_{2n}\Cbb$
and $\mathfrak{sp}_{2n}\Cbb$ can be analyzed in a similar
fashion  using the formulas from chapter 5 of \cite{k6494}. However,
due to the increase in complexity of the formulas over those
for $\Sl{n}$, the proofs become much more difficult and tedious.

To proceed, 
let $\mathfrak{D}$ denote the set of diagonal $n\times n$ complex
matricies. Then 
\eqn{slnCartan}{
\h = \{\, H \in \mathfrak{D} \, | \, \text{trace}(H) = 0 \, \}
}
is a Cartan subalgebra for $\Sl{n}$. Define $\epsilon_{j} \in \mathfrak{D}^{*}$
by
\eqn{epsilon}{
 \epsilon_{j}( \diag(H_{1},H_{2},\ldots,H_{n})) = H_{j} \: . 
} 
The set of roots determined by $\h$ is $R = \{\: \epsilon_{i} - \epsilon_{j}\: | \: 1\leq i,j \leq n \quad
i \neq j \: \}$ and  
$\Delta = \{\: \epsilon_{i} - \epsilon_{j} \: | \: j = 1,2,\ldots , n-1\: \}$ is
a base for $R$.
Suppose $\Lo$ is the $A_{1}$-vector determined by the partition $\db = (d_{1},
d_{2},\ldots, d_{k})$ according to the formulas \eqref{Ablock1} and \eqref{Ablock2}.
 
\begin{lem}\label{regA1lem2} \mnote{regA1lem2]}
If there exists $r,s \in \{1,2,\ldots,k\}$ with $r<s$ such that
$d_{r}$ and $d_{s}$ are both even, then $S_{\lambda}$ is not
a $\Pi$-system.
\end{lem}
\begin{proof}
Since $d_{r}$ and $d_{s}$ are both even, and $r < s$,
it follows that $d_{r} \geq d_{s} \geq 2$. Let $\hat{d}_{j} := \sum_{i=1}^{j-1} d_{i}$,
$I = \hat{d}_{r}+d_{r}/2$, and $J= \hat{d}_{s}+d_{s}/2$. Then it is not
difficult to verify that
$\epsilon_{I} - \epsilon_{I+1}$ and $\epsilon_{I} - \epsilon_{J+1}$ 
are in $S_{\lambda}$. But $(\epsilon_{I} - \epsilon_{I+1}) - 
(\epsilon_{I} - \epsilon_{J+1}) = \epsilon_{J+1} - \epsilon_{I+1} \in R$
and hence $S_{\lambda}$ is not a $\Pi$-system by definition.
\end{proof}

\begin{lem}\label{regA1lem3} \mnote{regA1lem3]}
If there exists $r,s \in \{1,2,\ldots,k\}$ with $r<s$ such that
$d_{r}$ and $d_{s}$ are both odd and $d_{r} > 1$, then $S_{\lambda}$ is not
a $\Pi$-system.
\end{lem}
\begin{proof}
Since $r < s$, $d_{r}$ and $d_{s}$ are odd, and $d_{r}>1$, we must
have  $d_{r}\geq 3$ and $d_{r} \geq d_{s}$. Let
$I = \hat{d}_{r}+(d_{r}-1)/2$ and $J= \hat{d}_{s}+(d_{s}-1)/2$ with
$\hat{d}_{j}$ defined as in lemma \ref{regA1lem2}. Then it is easy
to show that $\epsilon_{I} - \epsilon_{I+1}$ and $\epsilon_{I} - \epsilon_{J+1}$ 
are in $S_{\lambda}$. But then
$(\epsilon_{I} - \epsilon_{I+1}) -
(\epsilon_{I} - \epsilon_{J+1}) = \epsilon_{J+1} - \epsilon_{I+1} \in R$
and hence $S_{\lambda}$ is not a $\Pi$-system by definition.
\end{proof}

From these two lemmas it is clear that the only possibilities left
for a partition $\db$ to give rise to a $S_{\lambda}$ that
is a $\Pi$-system is if $\db$ satisfies $(i)$, $(ii)$, or $(iii)$
from the statement of the theorem. It can be verified that each
of these leads to a $S_{\lambda}$ that is a $\Pi$-system. We will only
verify the case $(i)$. Now, straightforward computation 
shows that $S_{\lambda} = \{\: \epsilon_{i} - \epsilon_{i+1} \: | \:
i = 1,\ldots,d_{1}-1\}\cup\{\: \epsilon_{d_{1}+i}-\epsilon_{d_{1}+i+1}
\: | \: i = 1,\ldots,d_{2}-1 \: \}$.
Thus $S_{\lambda} \subset \Delta$ which implies that $S_{\lambda}$ is
a $\Pi$-system.
\end{proof}

\begin{cor}\label{regA1thm2} \mnote{regA1thm2]} 
For the simple algebras $B_{\ell}$, $C_{\ell}$, $D_{\ell}$,
$G_{2}$, $F_{4}$, $E_{6}$, $E_{7}$, $E_{8}$, and
$A_{\ell'}$ with $\ell'$ odd, the only 
regular $A_{1}$-vector is the the principal one.
For $\ell'$ even, the regular $A_{1}$-vectors
are classified by partitions of the form
$\db = (\ell'+1-k,k)$ for $k=1,2,\ldots,\ell'/2$.
\end{cor}
\begin{proof} The statements concerning the classical simple
algebras $A_{\ell}$, $B_{\ell}$, $C_{\ell}$, and $D_{\ell}$
can be verified using theorem \ref{regA1thm2} 
and the formulas from chapter 5 of \cite{k6494}. For the
the exceptional algebras $G_{2}$, $F_{4}$, $E_{6}$, $E_{7}$, and 
$E_{8}$, the conclusion follows immediately
from theorem \ref{regA1thm1}.
\end{proof}

We know from lemma \ref{rA1lem} that
$\Av \cap  \overline{\Wc}
= \rAv \cap \overline{\Wc}$. But it is also clear from
theorem \ref{regA1thm1} and
the discussion in section \ref{sl2} that if
we let $\mathcal{M}_{\Pi} := 
\{\, \Lo \in \rAv \cap \overline{\Wc} \,|\, \text{
$S_{\lambda}$ is a $\Pi$-system}\,\}$ then $|\mathcal{M}_{\Pi}|$$\ll$ $|\Av \cap \overline{\Wc}|$.
Therefore $|\mathcal{M}_{\Pi}|$ $\ll$ $|\rAv \cap \overline{\Wc}|$
and we conclude that the regular models are rare.

\sect{chgb}{Global behavior}

In the papers \cite{eymg,eymgg}, we established that 
the EYM equations are locally solvable near the origin $r=0$
and a black hole horizon $r=r_{H}$. If any of these local solutions could
be continued out to $r=\infty$, we would
like to know its behavior. 
Knowing the global behavior is important
for two reasons. The first is that numerical solutions
can be constructed much more efficiently when
one knows what to expect. The second
is that we believe that these global estimates will
be necessary in proving the existence of global
solutions as was the case when $G=SU(2)$ \cite{k5157,k5201,k5428} and more recently
$G=SU(3)$  \cite{k6645}.

Suppose that $\{\Lp(r),m(r)\}$ is a bounded solution to
\eqref{feq1}, \eqref{feq3}, and \eqref{feq4} in
a neighborhood of $r=r_{*}$ where $r_{*}=0$ or
$r_{*}=r_{H} > 0$.
We are interested in the local solutions that
can be continued out to $r=\infty$ with
$N(r) > 0$ for $r> r_{*}$. 
For the moment we will assume that there
exists a $r_{0} > r_{*}$ so that the
conditions
\lgath{gbidata}{
N(r_{0}) < 1, \quad \|\Lp(r)\| \leq \frac{1}{\sqrt{2}}\|\Lo\| \; ,
\label{gbidata1}\\
\intertext{and}
[\Lp'(r_{0}),\Lm(r_{0})]+[\Lm'(r_{0}),\Lp(r_{0})] = 0 \label{gbidata2},
}
are satisfied.
At the end of section \ref{gb} we will show that all local solutions
that can be continued out to $r=\infty$ with
$N(r) > 0$ for $r>r_{*}$ will necessarily have
to satisfy these conditions.
Before we state the main theorem that characterizes the global behavior,
we first need to introduce a technical condition. The space
$V_{2}$ (see \eqref{V2Slam}) is uniquely determined by the choice of $\Lo$ in  
$\rAv \cap\overline{\Wc}$. Therefore the bilinear form 
\leqn{Bform}{
B : V_{2}\times V_{2} \longrightarrow V_{2} \: : \: (X,Y) \longmapsto
[X,c(Y)] \: .
}
depends implicitly on $\Lo$. Our results require that $\Lo$ is
chosen so that the following coercive condition is satisfied
\leqn{cc}{
\frac{4}{\norm{\Lo}^{2}} \leq
\inf_{X\in V_{2}\backslash \{0\}} \frac{\norm{B(X,X)}^{2}}{\norm{X}^{4}}\: .
}
We show in the next section that there exists $\Lo$ in  $\rAv \cap\overline{\Wc}$
for which the  inequality \eqref{cc} is satisfied. In fact we have
some evidence that \eqref{cc} is satisfied for all $\Lo$ in
$\rAv \cap\overline{\Wc}$ although we have no proof of this
fact. 

To state our main result we must first recall from \cite{eymg} that 
given a vector $\Omega_{+} \in V_{2}$ such that
$\{\Lo,\Omega_{+},\Omega_{-}:=-c(\Omega_{+})\}$ forms a real
standard triple then
the $\Rbb$-linear operator $\mathbf{\text{A}}:\g\rightarrow \g$ 
defined by
\eqn{Aop}{
\mathbf{\text{A}} := \half \ad(\Omega_{+}) \circ \left(
\ad(\Omega_{-}) + \ad(\Omega_{+})\circ c\right)
}
preserves the subspace $V_{2}$ and is diagonalizable. The
space $V_{2}$ then decomposes into
\eqn{Eplus}{
V_{2} = E_{0}\oplus E_{+}
} 
where $E_{0} := \ker \mathbf{\text{A}}\restr {V_{2}}$ and
$E_{+}$ is the direct sum of all the eigenspaces with
positive eigenvalues.

We now state our main result:
\begin{thm} \label{gbthm21} \mnote{[gbthm21]}
Suppose $\Lo \in \rAv \cap \overline{\Wc}$ is such that the inequality \eqref{cc} is
satisfied. If $\{\Lp(r),m(r)\}$ is a solution to equations \eqref{feq1} and \eqref{feq3}
defined on $[r_{0},\infty)$ $(r_{0} > 0)$ that satisfies
\eqn{gbidataA}{
N(r_{0}) < 1 \, , \quad \norm{\Lp(r_{0})} \leq \frac{1}{\sqrt{2}}\norm{\Lo}\, , \quad
[\Lp'(r_{0}),\Lm(r_{0})] + [\Lm'(r_{0}),\Lp(r_{0})] = 0 \, ,
}
at the point $r_{0}$ and 
\eqn{Ngtz}{
\text{$N(r) > 0$ for all $ r \geq r_{0}$,}
}
then
\begin{itemize}
\item[(i)]  \hspace{0.1cm} there exist a $m_{\infty} > 0$ such that $m(r) \rightarrow m_{\infty}$
as $r \rightarrow \infty$,
\item[(ii)]  $0 < N(r) < 1$ for all $r \geq r_{0}$,
\item[(iii)] equation \eqref{feq4} is automatically satisfied for all $r \geq r_{0}$,
\item[(iv)]  equation  \eqref{feq2} can be integrated to obtain $S(r)$ and $S(r_{0})$ can
be chosen so that  $S(r) \rightarrow 1$ as $r \rightarrow \infty$,
\item[(v)]  $\norm{\Lp(r)} \leq \norm{\Lo}/\sqrt{2}$ for all $r \geq r_{0}$,
\item[(vi)]  $r\Lp'(r) \rightarrow 0$  and $\norm{\Lp(r)-\Ff^{\times}} \rightarrow 0$
as $r\rightarrow \infty$,
\end{itemize}
where 
\eqn{Fxset}{
\Ff^{\times} := \{\: X \in V_{2}\setminus\{0\} 
\: | \: [[c(X),X],X] = 2 X\: \} .
}

Moreover if $S_{\lambda}$ is a $\Pi$-system then $\Lp(r) \in E_{+}$ for all $r
\geq r_{0}$ and $\lim_{r\rightarrow \infty} \Lp(r) = \Op^{\infty}$ for some
$\Op^{\infty} \in \Ff^{\times}\cap E_{+}$.
\end{thm}
When $\Lo$ is such that $S_{\lambda}$ is a $\Pi$-system, this theorem
is a natural generalization of the $SU(2)$ results.
However, if $S_{\lambda}$ is not a $\Pi$-system, then there is a possibility
for a new type of behavior as $(vi)$ leaves open the possibility that
$\Lp(r)$ does not actually approach a limit as $r\rightarrow \infty$. 
The reason that this possibility exists is that when $S_{\lambda}$ is
not a $\Pi$-system $\Ff^{\times}$ forms a 
$|S_{\lambda}|$-dimensional real variety. 
On the 
other hand, the existence of the limit $\Lp(r)$ as $r \rightarrow \infty$
when $S_{\lambda}$ is a $\Pi$-system 
is due to the fact that the set  $\Ff^{\times}\cap E_{+}$ is discrete ( see
lemma \ref{coerlem1a}).

From the definition
of $\Ff^{\times}$, it is clear that every $X_{+} \in \Ff^{\times}$
determines a real standard triple $\{X_{0},X_{-},X_{+}\}$
where $X_{-} := -c(X_{+})$ and $X_{0} := [X_{+},X_{-}]$.
For a given\mnote{was 'As'} $\Lo \in \rAv \cap \overline{\Wc}$, we know that there exist
an $\Op \in V_{2}$ such that $\{\Lo,\Op,\Om\}$ ($\Om:=-c(\Op)$) 
is a real standard triple.
Let 
\eqn{nmc}{
\Ef := \{\, \Op \in V_{2}\setminus\{0\}\, |\, 
[c(\Op),\Op] = \Lo \,\}.
}
Then $\Ef \subset \Ff^{\times}$, and it follows from \eqref{tYMcharge1}
that the magnetic charge $Q_{M} \rightarrow 0$ as
$r\rightarrow \infty$ if and only if $\|\Lp(r)-\Ef\| \rightarrow 0$
as $r\rightarrow \infty$. Therefore $\Ff^{\times}\setminus \Ef$ characterizes
the asymptotic values of $\Lp(r)$ for which the magnetic charge does
not vanish. If $G=SU(2)$ then $\Ef=\Ff^{\times}$ and so we recover
the known fact that the global solutions cannot have any magnetic
charge. For $G\neq SU(2)$, in general $\Ef$ is a proper
subset of $\Ff^{\times}$
and hence there exist a possibility for solutions with magnetic
charge. As we mentioned earlier in section \ref{restLo}, 
purely magnetic black hole
solutions with nonzero magnetic charge  have been found numerically \cite{k6174}. 
However, it is not clear if solitons with nonzero magnetic charge exist.
No numerical solutions of this type have been found.
Assuming that $\lim_{r\rightarrow \infty}\Lp(r)$ exists, the
initial value problem at $r=\infty$ may provide some insight.
For $\g=\Sl{n}$ and $\Lo$ principal, 
the possiblility of magnetic charge has been studied
in \cite{hka28}.
To describe these results, we first expand $\Lp(r)$ as
(see \eqref{Lpexp})
\eqn{Lpexp1}{
\Lp(r) = \sum_{\alpha\in S_{\lambda}} w_{\alpha}(r) \eb_{\alpha} \; ,
}
where the $w_{\alpha}$ are real valued functions. We note this
expansion is possible since $\Lo$ is regular.
If the magnetic charge does not vanish, then it can be shown that
$\lim_{r\rightarrow \infty} w_{\alpha}(r) = 0$ 
for some of the $\alpha \in S_{\lambda}$. Assuming analyticity of
the solution about $r=\infty$, the power series expansion then
shows that $w_{\alpha} = 0$ for $r$ near $r=\infty$. We expect
, although we have no proof, that $w_{\alpha}=0$ near $r=\infty$ actually implies that
$w_{\alpha} = 0$ for all $r$. For black hole solutions this
is not a problem. In fact the magnetically charged  black hole solutions of
\cite{k6174} were found by setting $w_{\alpha} = 0$ for
certain $\alpha \in S_{\lambda}$. But for
solitons,  $w_{\alpha} = 0$ for any $\alpha \in S_{\lambda}$
is not compatible
with the boundary conditions at $r=0$. This may explain why
no magnetically charged solitons have been found.
Our analysis of the initial value problem at $r=\infty$
\cite{eymg,eymgg}
has been done under the assumption that
$\lim_{r\rightarrow \infty}\Lp(r) \in \Ef$. In view of the
above discussion, it would be desirable to generalize
the existence and uniqueness proof at $r=\infty$ to
allow for $\lim_{r\rightarrow \infty}\Lp(r)$ in
$\Ff^{\times}$.

It was observed in \cite{eymgg}
that the existence proof for the gauge group
$SU(2)$ can be used to imply the existence of global solutions 
for any compact gauge
group and any generator $\Lo$ in $\rAv \cap\overline{\Wc}$.  
This of course leaves open the question of what are all the possible
global solutions. The rest of this article will be spent proving
theorem  \ref{gbthm21}.

\sect{coer}{A coercive condition}

In this section we  show that there exist 
$\Lo$ in $\rAv \cap\overline{\Wc}$ so that
\eqref{cc} is satisfied. To start,  we first derive an inequality that
is equivalent to \eqref{cc} but easier to work with.
Let 
\eqn{sph}{
S(V_{2}) := \{\: Y\in V_{2} \: |\: \norm{Y}=1\:\}
}
and define 
\eqn{Jset}{
\mathcal{J} := \{\: X \in S(V_{2}) \: | \: [[c(X),X],X] = \norm{[X,c(X)]}^{2} X
\: \}
}

\begin{lem} \label{coerlem1} \mnote{[coerlem1]}
\eqn{coerlem1.1}{
\inf_{X\in \mathcal{J}} \norm{B(X,X)}^{2} = 
\inf_{X\in V_{2}\backslash \{0\}} \frac{\norm{B(X,X)}^{2}}{\norm{X}^{4}}
}
\end{lem}
\begin{proof}
Define
\eqn{coerlem1.2}{
Q(Y) := B(Y,Y)^{2} \: ,
}
and let $\mathcal{C}$ denote the set of critical points
of $Q\restr{S(V_{2})}$. Then it is clear that 
\eqn{coerlem1.3}{
\inf_{X\in \mathcal{C}} \norm{B(X,X)}^{2} =               
\inf_{X\in V_{2}\backslash \{0\}} \frac{\norm{B(X,X)}^{2}}{\norm{X}^{4}}\: .
}
Therefore to prove the theorem we need to show that $\mathcal{J}=\mathcal{C}$.
So suppose $X$ is a critical point of $Q\restr{S(V_{2})}$ and let
$f(Y) := \norm{Y}^{4}$. By the method of Lagrange multipliers
there exists a $\beta \in \mathbb{R}$ such
that
\eqn{coerlem1.4}{
DQ(X) = \beta Df(X) \: .
}
Straightforward calculation shows that
\eqn{coerlem1.5}{
Df(Z)\cdot Y = 4\norm{Z}^{2}\rip{Z}{Y} \AND
DQ(Z)\cdot Y = -4\rip{[[Z,c(Z)],Z]}{Y} \: .
}
Therefore $X$ must satisfy
\leqn{coerlem1.6}{
\norm{X} = 1 \AND [[c(X),X],X] = \beta X \: .
}
Taking the norm on both sides of $[[c(X),X],X] = \beta X$
and using $\norm{X} = 1$ yields 
\leqn{coerlem1.7}{
\beta = \rip{[[c(X),X],X]}{X} = \norm{[c(X),X]}^{2} .
}
Let $X_{+} := X$, $X_{-} := -c(X)$,
and $X_{0} := [X_{+},X_{-}]$. Then
$[X_{0},X_{\pm}] = \pm \norm{X_{0}}^{2} X_{\pm}$
by \eqref{coerlem1.6} and
\eqref{coerlem1.7}.
This proves that $\mathcal{C} \subset  \mathcal{J}$. The reverse
inclusion is straightforward to verify.
\end{proof}

Define
\leqn{fixpts1}{
\Ff := 
\{\: X \in V_{2} \: | \: [[c(X),X],X] = 2 X
\: \}
}
At this point we will prove a result about the structure of
$\Ff$ that will be required later on. This result will not be
used in this section. 
\begin{lem} \label{coerlem1a} \mnote{[coerlem1a]}
If $S_{\lambda}$ is a $\Pi$-system, then $\Ff\cap E_{+}$
is a discrete set.
\end{lem}
\begin{proof}
Proposition 4 in \cite{eymg}  shows that
$E_{+} = \sum_{j=1}^{\ell}\Rbb \eb_{\alpha_{j}}$, so we can
expand $X_{+} \in E_{+}$ as
\eqn{coerlem1a.2}{
X_{+} = \sum_{j=1}^{\ell} x_{j} \eb_{\alpha_{j}}
} 
where $x_{j} \in \Rbb$. So 
\eqn{coerlem1a.3}{
X_{-} := -c(X_{+}) = \sum_{j=1}^{\ell} x_{j}\eb_{-\alpha_{j}} 
}
and hence
\eqn{coerlem1a.4}{
X_{0} := [X_{+},X_{-}]  = \sum_{j=1}^{\ell} \sum_{k=1}^{\ell} x_{j}x_{k}
[\eb_{\alpha_{j}},\eb_{-\alpha_{k}}] =
\sum_{j=1}^{\ell} x_{j}^{2}\hb_{\alpha_{j}} 
}
as $[\eb_{\alpha_{j}},\eb_{-\alpha_{k}}] = \delta_{jk}\hb_{\alpha_{j}}$.
Using $[\hb_{\alpha_{j}},\eb_{\pm\alpha_{k}}] = \pm C_{kj} \eb_{\pm\alpha_{k}}$
where $C_{kj}$ is the Cartan matrix of $\g_{\lambda}$, we get
\eqn{coerlem1a.5}{
[X_{0},X_{\pm}]  \mp 2 X_{\pm} =
\pm\sum_{k=1}^{\ell}\left(\sum_{j=1}^{\ell}C_{kj}x_{j}^{2}-2\right)x_{k}
 \eb_{\pm\alpha_{k}} \; .
}
Because the vectors $\eb_{\pm\alpha_{j}}$ are linearly independent, it
is clear that $X_{+} \in \Ff\cap E_{+}$ if and only if
\eqn{coerlem1a.6}{
\left(\sum_{j=1}^{\ell}C_{kj}x_{j}^{2}- 2\right)
x_{k}=0 \quad \text{ for $k=1,2,\ldots
\ell$.}
}
Using the invertibility of the Cartan matrix $C$, 
the above equation can be solved to give 
\eqn{coerlem1a.7}{
x_{k}=0\quad \text{or} \quad x_{k} = \pm \left(2\sum_{j=1}^{\ell}
(C^{-1})_{kj} \right)^{\frac{1}{2}} \quad k=1,2,\ldots, \ell \; .
}
This solution set is obviously finite and therefore the proof is
complete.
\end{proof}
Define
\leqn{fixpts2}{
\Ff^{\times} := \Ff\backslash \{0\} .
}
\begin{lem} \label{coerlem2} \mnote{[coerlem2]}
\eqn{coerlem2.1}{
\inf_{X\in \mathcal{J}} \norm{B(X,X)}^{2} =
\inf_{X\in \Ff^{\times}} \frac{4}{\norm{B(X,X)}^{2}} 
}
\end{lem}
\begin{proof}
Suppose $X_{+} \in \mathcal{J}$.  Let $X_{-} := -c(X_{+})$ and $X_{0} := [X_{+},X_{-}]$. Then
\leqn{coerlem2.2}{
2\norm{X_{+}}^{2} = \rip{2 X_{+}}{X_{+}}
=\rip{[X_{0},X_{+}]}{X_{+}} = \rip{X_{0}}{[c(X_{+}),X_{+}]} = \norm{X_{0}}^{2} \: .
}
Also note that if $X\in V_{2}$ and $[c(X),X] = 0$ then
\eqn{coerlem2.3}{
0 = \rip{[c(X),X]}{\Lo} = \rip{X}{[X,\Lo]} = -2\norm{X}^{2} \: .
}
Therefore
\leqn{coerlem2.4}{
\text{if $X\in V_{2}$ then $[c(X),X]=0$ if and only if $X=0$.}
}

Define a map
\eqn{coerlem2.5}{ 
\hat{} : \mathcal{J} \longrightarrow \Ff^{\times}
}
by
\eqn{coerlem2.6}{
\quad X_{+} \longmapsto \hat{X}_{+} := \frac{\sqrt{2}}{\norm{[c(X_{+}),X_{+}}} \
X_{+}\, .
}
Then using \eqref{coerlem2.2} and \eqref{coerlem2.4}, it is straightforward to 
verify that the above map is well defined and bijective. The proof now
follows since
\eqn{coerlem2.7}{
\|B(X_{+},X_{+})\|^{2} = \frac{\|B(\hat{X}_{+},\hat{X}_{+})\|^{2}} 
{\|\hat{X}_{+}\|^{4}} \: ,
} 
and 
\eqn{coerlem2.8}{
\frac{\|B(\hat{X}_{+},\hat{X}_{+})\|^{2}} 
{\|\hat{X}_{+}\|^{4}} = \frac{4}{\|B(\hat{X}_{+},\hat{X}_{+})\|^{2}} 
}
by \eqref{coerlem2.2} and the fact that $\|B(\hat{X}_{+},\hat{X}_{+})\|
= \|X_{0}\|$.
\end{proof}

The above two lemmas show that the coercive condition \eqref{cc} is
equivalent to
\leqn{cc1}{
\frac{4}{\norm{\Lo}^{2}} \leq \inf_{X\in \Ff^{\times}} \frac{4}{\norm{B(X,X)}^{2}}\; .
}
We will now show that there exist generators $\Lo$ in
$\rAv \cap\overline{\Wc}$ that satisfy
the inequality \eqref{cc}.
 
\begin{thm} \label{coerthm3} \mnote{[coerthm3]}
If $S_{\lambda}$ is a $\Pi$-system then the inequality \eqref{cc} is satisfied.
\end{thm}
\begin{proof}
Since $S_{\lambda}$ is a $\Pi$-system, the discussion in section
\ref{regA1} shows that we can without loss of generality assume
that $\Lo$ is a principal $A_{1}$-vector.
Note that if $X_{+} \in \Ff$ then $X_{0} := [X_{+},X_{-}] \in
\Av$ where $X_{-} := -c(X_{+})$. 
Also note that since $X_{0}$ satisfies $c(X_{0})=-c(X_{0})$ it follows
from the definition of $\norm{\cdot}$ and $B$ that
$\Kill{X_{0}}{X_{0}} = \norm{B(X_{+},X_{+})}$. Therefore 
\leqn{coerthm3.2}{
\inf_{X\in \Av} \frac{4}{\Kill{X}{X}} \leq \inf_{X\in 
\Ff^{\times}} \frac{4}{\norm{B(X,X)}^{2}}\; ,
}

as it can be easily shown that $\Kill{X}{X}\in \mathbb{R}$ for all
$X \in \Av$. 

\begin{lem} \label{coerlem4} \mnote{[coerlem4]}
If $Y_{0} \in \mathcal{A}_{1}^{\text{\emph{v}}}$ then $\Kill{Y_{0}}{Y_{0}} \leq \norm{\Lo}^{2}$.
\end{lem}
\begin{proof}
Since $\Lo$ is principal, there exists a base $\Delta$ such that
\leqn{coerlem4.1}{
\text{$\alpha(\Lo) = 2$ for all $\alpha \in \Delta$.} 
}
Also there exists an automorphism $\phi$ of $\g$ such that 
\leqn{coerlem4.2}{
\text{$\alpha(\phi(Y_{0})) = 0$, $1$, or $2$ for every $\alpha \in \Delta$.}
}
Now
\eqn{coerlem4.3}{
\Lo = \sum_{\alpha\in\Delta} \lambda_{\alpha} \hb_{\alpha} \AND
\phi(Y_{0}) = \sum_{\alpha\in\Delta} y_{\alpha} \hb_{\alpha}
}
where 
\eqn{coerlem4.4}{
\lambda_{\alpha} = 2 \sum_{\beta\in\Delta} (C^{-1})_{\alpha\beta}\; , \quad 
y_{\alpha} = \sum_{\beta\in\Delta} (C^{-1})_{\alpha\beta}\:
\beta(\phi(Y_{0})) 
}
and $C^{-1}$ is the inverse of the Cartan matrix $C=(\langle \alpha , 
\beta \rangle)$. Using the above expansions, it is easy to show that 
\leqn{coerlem4.5}{
\Kill{\Lo}{\Lo} = 2 \sum_{\alpha , \beta \in \Delta}
\frac{4}{|\alpha|^{2}} (C^{-1})_{\alpha\beta} \; ,
}
and
\leqn{coerlem4.6}{
\Kill{Y_{0}}{Y_{0}} = 2 \sum_{\alpha , \beta \in \Delta}
\frac{1}{|\alpha|^{2}} \alpha(\phi(Y_{0}))\: (C^{-1})_{\alpha\beta}
\: \beta(\phi(Y_{0})) \; .
}
But $ (C^{-1})_{\alpha\beta} \geq 0$ for all $\alpha$, $\beta\in \Delta$.
Therefore $\Kill{\phi(Y_{0})}{\phi(Y_{0})} \leq
\Kill{\Lo}{\Lo}$ by \eqref{coerlem4.2}, \eqref{coerlem4.5},
and \eqref{coerlem4.6}.  Finally, observe that
$\Kill{\phi(Y_{0})}{\phi(Y_{0})} = \Kill{Y_{0}}{Y_{0}}$ and
$\Kill{\Lo}{\Lo} = \norm{\Lo}^{2}$ since
$\phi$ is an automorphism and
$c(\Lo) = -\Lo$. Therefore 
$\Kill{Y_{0}}{Y_{0}} \leq \norm{\Lo}^{2}$ and the proof is complete. 
\end{proof}
From this lemma and \eqref{coerthm3.2}, we see that the inequality
\eqref{cc1} is satisfied. By the above results this implies that \eqref{cc}
is also satisfied. 
\end{proof}

Since we now know that there exists $\Lo$ in $\rAv \cap \overline{\Wc}$
such that the inequality \eqref{cc} is satisfied, it would  be 
desirable to determine exactly which $\Lo$ satisfy \eqref{cc}.
In general, this appears to be a difficult question. However, for
low dimensional Lie algebras, computations show
that every $\Lo$ in $\rAv \cap \overline{\Wc}$ satisfies \eqref{cc}.
This gives some evidence to our belief that \eqref{cc} is
always satisfied. If this were the case, then our later proofs
that rely on \eqref{cc} would be general.

\sect{aym}{Asymptotic Yang-Mills equations}

The flat space spherically symmetric Yang-Mills equations can be written as
\leqn{fym}{
\ddot{\Lambda}_{+} - \dot{\Lambda}_{+} + \Fc = 0
}
where $\dot{(\cdot)} = \frac{d(\cdot)}{d\tau}$ and $\tau = \ln(r)$. However,
for the purpose of this section we will consider equation \eqref{fym} in its
own right, and let $\tau$ denote a  parameter that is 
not necessarily related to the radial coordinate $r$.  
We will be interested in $\tau \rightarrow \infty$ behavior of
bounded solutions to equations of the form
\leqn{afym}{
\ddot{\Lambda}_{+} - \dot{\Lambda}_{+} + \Fc = \delta(\tau)\dot{\Lambda}_{+}
\: ,
}
where $\delta$ is any $C^{1}$ function that satisfies
\leqn{delta}{
\lim_{\tau \rightarrow \infty} \delta(\tau) = 0 \: .
}
To determine this behavior, we use the
results of Markus \cite{k6678} concerning the
long time behavior of solutions to asymptotically
autonomous differential equations. See also
\cite{k6679,k6590}. To describe
these results we first recall that a nonautonomous
system of differential equations in $\Rbb^{N}$
\leqn{Mar1}{
\dot{x}(\tau) = h(\tau,x(\tau))
}
is said to be \emph{asymptotically autonomous}
with \emph{limit equation}
\leqn{Mar2}{
\dot{y}(\tau)= g(y(\tau))
}
if
\eqn{Mar3}{
\text{$h(\tau,x) \rightarrow g(x)$ as $\tau\rightarrow \infty$ 
uniformly on compact subsets of $\Rbb^{N}$.}
}
We note that the maps $h$ and $g$ are assumed to be
continuous and locally Lipschitz on $\Rbb^{N}$.
The $\omega$-limit set $\omega(\tau_{0},x_{0})$ of
a bounded solutions $x(\tau)$ to \eqref{Mar1}
on $[\tau_{0},\infty)$ satisfying $x(\tau_{0})=x_{0}$
is defined by
\eqn{ols}{
\omega(\tau_{0},x_{0}) = \{\, y \, |\, \text{$y=\lim_{j\rightarrow
\infty}x(\tau_{j})$ for some sequence $\tau_{j} \rightarrow \infty$}\,\}
\, .
}
The fundamental result of Markus is:
\begin{thm} \label{Markus} \mnote{[Markus]}
The $\omega$-limit set  $\omega(\tau_{0},x_{0})$ of
a bounded solution $x(\tau)$ to \eqref{Mar1}
on $[\tau_{0},\infty)$ satisfying $x(\tau_{0})=x_{0}$
is nonempty, compact, and connected. Moreover,
\eqn{Mar4}{
\text{$\text{\emph{dist}}(x(\tau),\omega(\tau_{0},x_{0})) 
\rightarrow 0$ as $\tau \rightarrow\infty$} \, , 
}
and $\omega(\tau_{0},x_{0})$ is invariant under
\eqref{Mar2}.
\end{thm}

Define maps

\lgath{Ffgmaps}{
\Fch : V_{2} \longrightarrow V_{2} \: : \: 
X \longmapsto \frac{1}{2}[\Lo+[X,c(X)],X] \, , \label{Fmap} \\
f : \mathbb{R}\times V_{2}\times V_{2} \longrightarrow V_{2}\times V_{2}
\: :\: (\tau,X_{1},X_{2}) \longmapsto (X_{2},X_{2}-\Fch(X_{1})+
\delta(\tau)X_{2}) \, , \label{fmap} \\
\intertext{and}
g : V_{2}\times V_{2} \longrightarrow V_{2}\times V_{2}
\: :\: (\tau,X_{1},X_{2}) \longmapsto (X_{2},X_{2}-\Fch(X_{1})
) \, . \label{gmap}
}
Using these maps we can write \eqref{fym} and \eqref{afym} in
first order form as
\leqn{fym1}{
(\dot{\Lambda}_{+},\dot{\Gamma}_{+}) = g(\Lp,\Gp)
}
and
\leqn{afym1}{
(\dot{\Lambda}_{+},\dot{\Gamma}_{+}) = f(\tau,\Lp,\Gp)\; ,
}
respectively.

The proof of the next proposition is straightforward
and left to the reader.
\begin{prop} \label{aymprop1} \mnote{[aymprop1]}
$f(\tau,X,Y) \longrightarrow g(X,Y)$ as $\tau \rightarrow \infty$ 
uniformly on compact subsets of $V_{2}\times V_{2}$. 
\end{prop}

This proposition shows that the nonautonomous system \eqref{afym1}
is asymptotically autonomous with limit equation \eqref{fym1}.

\begin{prop} \label{aymprop2} \mnote{[aymprop2]}
Suppose  $\mathbf{X}(\tau) = (X_{1}(\tau),X_{2}(\tau))$ is a bounded
solution to \eqref{afym1} that is defined for all $\tau \geq \tau_{0}$
and satisfies $\mathbf{X}(\tau_{0}) = \mathbf{X}_{0}$. 
Then
\begin{itemize}
\item[(i)]  $\qquad \omega(\tau_{0},\mathbf{X}_{0})$ is non-empty, compact and
connected, 
\item[(ii)] $\qquad \norm{\mathbf{X}(\tau)-\omega(\tau_{0},\mathbf{X}_{0})} 
\rightarrow 0$ as $\tau \rightarrow \infty$,
\item[(iii)] $\qquad  \omega(\tau_{0},\mathbf{X}_{0})$ 
is invariant under \eqref{fym1}.
\end{itemize}
\end{prop}
\begin{proof}
Follows directly from theorem \ref{Markus} by proposition
\ref{aymprop1}.
\end{proof}

Define 
\leqn{H}{
H : V_{2} \times V_{2} \longrightarrow \mathbb{R} 
\: : \: (X_{1},X_{2}) \longmapsto \frac{1}{2}\norm{X_{2}}^{2}- \frac{1}{2}
\|\Fh(X_{1})\|^{2}\: ,
}
where
\leqn{Fhat}{
\Fh(X) := \frac{i}{2}(\Lo + [X,c(X)]) \: .
}
\begin{prop} \label{aymprop3} \mnote{[aymprop3]}
If $\mathbf{X}(\tau) = (X_{1}(\tau),X_{2}(\tau))$ is a bounded
solution to \eqref{afym1}, then there exists a $\beta \in \mathbb{R}$
such that $H(\omega(\tau_{0},\mathbf{X}_{0})) = \beta$.
\end{prop}
\begin{proof}
Straightforward calculation using \eqref{cLo}, \eqref{Vn}, 
the properties \eqref{rip} of the inner product $\rip{\cdot}{\cdot}$, 
and \eqref{afym1} shows that
\eqn{aymprop3.1}{
(H(\mathbf{X}(\tau))\dot{\,} = \norm{X_{2}(\tau)}^{2}(1+\delta(\tau))\: .
}
But $\delta(\tau)\rightarrow 0$ as $\tau \rightarrow \infty$, which
shows that $(H(\mathbf{X}(\tau))\dot{\,} \geq 0$
for $\tau$ large enough. As $\mathbf{X}$ is bounded,
$\lim_{\tau\rightarrow \infty} 
H(\mathbf{X}(\tau))$
exists and we denote the limit by 
$\beta$. Therefore for any sequence $\tau_{k} 
\rightarrow \infty$, we also have 
$\lim_{k\rightarrow \infty} H(\mathbf{X}(\tau_{k})) = \beta$. By continuity of
$H$, we have $H(\lim_{k\rightarrow \infty}\mathbf{X}(\tau_{k})) = \beta$.
From the definition of $\omega(\tau_{0},\mathbf{X}_{0})$ it is clear
that $H(\omega(\tau_{0},\mathbf{X}_{0})) = \beta$.
\end{proof}

The fixed points of \eqref{fym1} are
\leqn{fixpts}{
\Ff\times \{0\}
}
where $\Ff$ was previously defined in \eqref{fixpts1}.

\begin{thm} \label{aymthm4} \mnote{[aymthm4]}
If $\mathbf{X}(\tau) = (X_{1}(\tau),X_{2}(\tau))$ is a bounded
solution to \eqref{afym1}, then
\begin{itemize}
\item[(i)] $\qquad \norm{X_{1}(\tau)-\Ff} \rightarrow 0$ as $\tau \rightarrow \infty$
\item[(ii)] $\qquad X_{2}(\tau) \rightarrow 0$ as $\tau \rightarrow \infty$
\end{itemize}
\end{thm}
\begin{proof}
Suppose $\mathbf{Y}_{0} = (Y_{1,0},Y_{2,0})  
\in \omega(\tau_{0},\mathbf{X}_{0})$.  Let $\mathbf{Y}(\tau)=(Y_{1}(\tau),Y_{2}(\tau))$ 
be a solution to \eqref{fym1} with $\mathbf{Y}(0)= \mathbf{Y}_{0}$. 
Then $\mathbf{Y}(\tau)  \in \omega(\tau_{0},\mathbf{X}_{0})$ and 
$H(\mathbf{Y}(\tau)) = \beta$ for all $\tau \geq 0$ by propositions 
\ref{aymprop2} and \ref{aymprop3}. Using 
\eqref{cLo}, \eqref{Vn},
the properties \eqref{rip} of the inner product $\rip{\cdot}{\cdot}$,
and \eqref{fym1}, it is not difficult to show that
$(H(\mathbf{Y}(\tau))\dot{\,} = \norm{Y_{2}(\tau)}^{2}$. Therefore we must
have $\norm{Y_{2}(\tau)}=0$ and hence $Y_{2}(\tau)=\dot{Y}_{2}(\tau)=0$.
It then follows form the differential equation \eqref{fym1} that 
$\dot{Y}_{1}(\tau)=0$ and $[c(Y_{1}(\tau)),Y_{1}(\tau)]=2Y_{1}(\tau)$.
Therefore, $\mathbf{Y}(\tau)=(Y_{0,1},0)$ and $[c(Y_{1,0}),Y_{1,0}]=2Y_{1,0}$.
This proves that $\omega(\tau_{0},\mathbf{X}_{0}) \subset \Ff\times \{0\}$.
The proof now follows easily since $\norm{\mathbf{X}(\tau)-\omega(\tau_{0},\mathbf{X}_{0})}
\rightarrow 0$ as $\tau \rightarrow \infty$ by proposition \eqref{aymprop2}.
\end{proof}
\begin{thm} \label{aymthm5} \mnote{[aymthm5]}
If $\mathbf{X}(\tau) = (X_{1}(\tau),X_{2}(\tau))$ is a non trivial
bounded solution to \eqref{afym1}, then
\begin{itemize}
\item[(i)] $\qquad \norm{X_{1}(\tau)-\Ff^{\times}} 
\rightarrow 0$ as $\tau \rightarrow \infty$
\item[(ii)] $\qquad X_{2}(\tau) \rightarrow 0$ as $\tau \rightarrow \infty$
\end{itemize}
\end{thm}
\begin{proof}
If $\lim_{\tau\rightarrow \infty} \mathbf{X}(\tau) \neq 0$ then we
are done by the above theorem. So assume that 
$\lim_{\tau\rightarrow \infty} \mathbf{X}(\tau) = 0$.
Let $W = V_{2}\times V_{2}$ and 
define a linear operator $\Tbb$ on $W$ by
\eqn{aymthm5.1}{
\Tbb(\mathbf{Z}) := \mathbf{D}g(0)\cdot \mathbf{Z} \: .
}
A short calculation shows that
\eqn{aymthm5.2}{
\Tbb = \begin{bmatrix}
 0 & \id \\
-\id & \id  
\end{bmatrix} \; ,
}
and  that $\Tbb$ has two distinct eigenvalues
$(1\pm i\sqrt{3})/2$ each with multiplicity $\dim_{\Rbb} V_{2}$.
Therefore there exists constants $K,\, \alpha > 0$ such that
\leqn{aymthm5.3}{
\big\lvert e^{-\tau\Tbb}\big\rvert \leq K e^{-\alpha \tau} \quad \forall \: \tau
\geq 0 \; .
}
Choose $l > 0$ so that
\leqn{aymthm5.4}{
l < \frac{\alpha}{K} \; .
}
Because $g(0)=0$ and $\mathbf{D}g(0)=0$ it can be shown
using appropriate smooth bump functions
that for any $\mu > 0$ there exists an $\epsilon > 0$ and a $C^{\infty}$ map 
$\hat{g} : W \rightarrow W$ such that
\lgath{aymthm5.5a}{
\norm{\hat{g}(\mathbf{Z}_{1})-\hat{g}(\mathbf{Z}_{2})}
\leq l
\norm{\mathbf{Z}_{1}-\mathbf{Z}_{2}} \quad \forall
\: \mathbf{Z}_{1},\mathbf{Z}_{2} \in W \: ,\label{aymthm5.5} \\
\norm{\hat{g}(\mathbf{Z} )} \leq \mu \quad \forall \mathbf{Z} \in W   
\label{aymthm5.5b}
}
and
\leqn{aymthm5.6}{
\hat{g}(\mathbf{Z}) = g(\mathbf{Z})-\Tbb(\mathbf{Z})  \quad \forall \:
 \mathbf{Z} \in B_{\epsilon}(W) \: .
}
Also because $\lim_{\tau\rightarrow \infty}\delta(\tau) = 0$,
there exists a 
$\tau_{0}$ and a $C^{\infty}$ function
$\hat{\delta}(\tau)$ such that
\leqn{aymthm5.7}{
|\hat{\delta}(\tau)| \leq l \quad \forall \tau \in \Rbb \: ,
}
and
\leqn{aymthm5.8}{
\hat{\delta}(\tau) = \delta(\tau) \quad \forall \tau \geq \tau_{0} \: .
}
Letting $\pr_{2} : W \rightarrow V_{2}$ denote projection onto the
second factor,  it is clear from \eqref{aymthm5.6} and
\eqref{aymthm5.8} that
\leqn{aymthm5.9}{
f(\tau,\mathbf{Z}) = \Tbb(\mathbf{Z}) + \hat{g}(\mathbf{Z}) +
\hat{\delta}(\tau)\pr_{2}(\mathbf{Z}) \quad
\forall \; \mathbf{Z} \in B_{\epsilon}(W), \: \tau \geq \tau_{0} \: .
}
Because $\lim_{\tau\rightarrow \infty} \mathbf{X}(\tau) = 0$, there
exists a $\tau_{1} > \tau_{0}$ such that $\mathbf{X}(\tau)
\in B_{\epsilon}(W)$ for all $\tau \geq \tau_{1}$. So
$\mathbf{X}(\tau)$ must be a solution to the differential
equation
\eqn{aymthm5.10}{
\dot{\mathbf{Y}} = \Tbb(\mathbf{Y}) + 
\hat{\delta}(\tau)\pr_{2}(\mathbf{Y}) + \hat{g}(\mathbf{Y})
}
for $\tau \geq \tau_{1}$ by \eqref{aymthm5.9}. Define
\leqn{aymthm5.10a}{
\Tbb(t) := \Tbb + \hat{\delta}(\tau)\pr_{2}
}
so that $Y$ satisfies 
\leqn{aymthm5.10b}{
\dot{\mathbf{Y}} = \Tbb(t)(\mathbf{Y}) + \hat{g}(\mathbf{Y}) \; .
}
Now $|\pr_{2}|\leq 1$ so $|\hat{\delta}(\tau)\pr_{2}| \leq l$ for all
$\tau \in \Rbb$ by \eqref{aymthm5.7}. Consequently,
\leqn{aymthm5.10c}{
\int_{\tau}^{\tau_{0}} |\hat{\delta}(s)\pr_{2}|ds \leq l(\tau_{0}-\tau) 
\quad \forall \tau_{0} \geq \tau\; .
}
Let $\Psi(\tau)$ be a fundamental matrix associated to $\Tbb(\tau)$. In
other words $\Psi(\tau)$ is an invertible matrix solution to 
\eqn{aymthm5.10da}{
\dot{\Psi}(\tau) = \Tbb(\tau)\Psi(\tau) \; .
}
It then follows by \eqref{aymthm5.3}, \eqref{aymthm5.4}, \eqref{aymthm5.10c},
and theorem 2.3 page 86 of \cite{k6680} that
$\Psi(\tau)$ satsifies
\leqn{aymthm5.10d}{
\lvert\Psi(\tau)\Psi(\tau_{0})^{-1}\rvert \leq e^{-\alpha(\tau_{0}-\tau)}
\quad \forall \; \tau_{0} \geq \tau \; .
}

The inequalities \eqref{aymthm5.5} and \eqref{aymthm5.5b}
guarantee that any solution of \eqref{aymthm5.10b} is defined for
all $\tau$. Let $\tilde{\mathbf{X}}(\tau)$ denote the unique global
solution to \eqref{aymthm5.10b} that satisfies 
$\tilde{\mathbf{X}}(\tau) = \mathbf{X}(\tau)$ for all $\tau \geq \tau_{1}$.
Since $\lim_{\tau\rightarrow \infty} \mathbf{X}(\tau) = 0$, 
$\tilde{\mathbf{X}}(\tau)$ is bounded on $[0,\infty)$. Notice, that
because $\hat{g}(0)=g(0)=0$, $Z(\tau) = 0$ is also a solution to 
\eqref{aymthm5.10b}. However,
a slight generalization of lemma 1.5, page 54 of
\cite{k6682} shows that any solution to \eqref{aymthm5.10b} bounded
on $[0,\infty)$ is unique by \eqref{aymthm5.4}, \eqref{aymthm5.5},
\eqref{aymthm5.5b}, and \eqref{aymthm5.10d}.  
Therefore, $\tilde{\mathbf{X}}(\tau) = 0$ for all $\tau \in \Rbb$ and
this implies that $\mathbf{X}(\tau)=0$ for $\tau \geq \tau_{1}$. But
$\mathbf{Z}(\tau) = 0$ is a solution to \eqref{afym1} and
so $\mathbf{X}(\tau)=0$ for all $\tau \in \Rbb$. This contradicts the assumption
that $\mathbf{X}(\tau)$ is a non-trivial solution to
\eqref{afym1}. Therefore  $\lim_{\tau\rightarrow \infty} \mathbf{X}(\tau) \neq 0$.
\end{proof}

\sect{gb}{Global estimates}

At the end of this section  we prove theorem \ref{gbthm21}.
However, we first need to prove a number of preliminary results.

\begin{prop} \label{gbprop1} \mnote{[gbprop1]}
If $\{\Lp(r),m(r)\}$ is a solution to equations \eqref{feq1} and \eqref{feq3}
defined on an interval $[r_{0},r_{1})$ $(r_{0} > 0)$ and that satisfies 
$N(r) > 0$ for all $ r \geq r_{0}$ and $[\Lp'(r_{0}),\Lm(r_{0})]
+ [\Lm'(r_{0}),\Lp(r_{0})] = 0$ then $\Lp$ also satisfies
\eqref{feq4}.
\end{prop}
\begin{proof}
Lemma 1 of \cite{eymgg}  shows that
$\gamma(r) = [\Lp(r),\Lm'(r)] + [\Lm(r),\Lp'(r)]$ satisfies
the differential equation
$\gamma' = -\frac{2}{r^{2}N}\left(m-\frac{1}{r}P\right)\gamma$. Integrating
this equation yields
\eqn{gbprop1.1}{
\gamma(r) = \gamma(r_{0})\exp\left(\int_{r_{0}}^{r}
-\frac{2}{s^{2}N}\left(m-\frac{1}{s}P\right)ds\right) \; .
}
But $\gamma(r_{0})=0$ by assumptions, hence $\gamma(r) = 0$ for all $r \geq r_{0}$.
\end{proof}

\begin{prop} \label{gbprop2} \mnote{[gbprop2]}
If $\{\Lp(r),m(r)\}$ is a solution to equations \eqref{feq1} and \eqref{feq3},
that satisfies $N(r_{0}) > 0$, $[\Lp(r_{0}),\Lm(r_{0})] = \Lo$, and 
$\Lp'(r_{0}) = 0$ for some $r_{0} > 0$, then
$\Lp(r) = \Lp(r_{0})$, $ m(r) = \frac{r_{0}}{2}(1-N(r_{0}))$, and
$S(r) = S(r_{0})$ for all $ r >  \max\{2m(r_{0}),0\}$.
\end{prop}
\begin{proof}
It is straightforward to check that if  
$N(r_{0}) > 0$, $[\Lp(r_{0}),\Lm(r_{0})] = \Lo$, and 
$\Lp'(r_{0}) = 0$, then $\Lp(r) := \Lp(r_{0})$ and 
$m(r) :=  \frac{r_{0}}{2}(1-N(r_{0}))$ solve \eqref{feq1}, \eqref{feq3},
and \eqref{feq4}. By standard uniqueness results for systems of
differential equations, this is the
only solution satisfying $N(r_{0}) > 0$, $[\Lp(r_{0}),\Lm(r_{0})] = \Lo$, and
$\Lp'(r_{0}) = 0$.
\end{proof}

The next two propositions generalize propositions 8 and 9 in 
\cite{k5428} which are valid for $G=SU(2)$.
. 
\begin{prop} \label{gbprop3} \mnote{[gbprop3]}
If $\{\Lp(r),m(r)\}$ is a solution to  equations \eqref{feq1} and \eqref{feq3},
on $[r_{0},r_{1})$ $(r_{0} > 0)$  and $0 < N(r_{0}) < 1$ then $N(r) < 1$ for all
$r\in [r_{0},r_{1})$.
\end{prop}
\begin{proof}
This can be proved in the exact same manner as when $G=SU(2)$. 
See \cite{k5428} proposition 8 for details. 
\end{proof}

\begin{prop} \label{gbprop4} \mnote{[gbprop4]}
If $\{\Lp(r),m(r)\}$ is a solution to equations \eqref{feq1} and \eqref{feq3},
on $[r_{0},r_{1})$ $(r_{0} > 0)$ with
$0 < \epsilon \leq N(r) < 1$ then there exists a $\delta > 0$ such
that the solutions exists and is analytic on $[r_{0},r_{1}+\delta)$. 
\end{prop}
\begin{proof}
First note that if the solution $\{\Lp(r),m(r)\}$ exists on some open interval 
$I \subset (0,\infty)$  on
which $N > 0$ then the Cauchy-Kowalevski theorem will
guarantee the solution will be analytic.
From standard theorems on differential equations, it follows
that the solution will continue to exist at $r=r_{1}$ unless
$N \rightarrow 0$ or one of the
variables $\{m,\Lp,\Lp'\}$ becomes unbounded as $r \rightarrow r_{1}$.
By assumption $N$ does not approach zero and
$ 0 < N(r) < 1$ implies that 
\leqn{gbprop4.1}{
0 < 2m(r) < r_{1} \quad \forall \; r \in [r_{0},r_{1}) \: .
}
Therefore we only need to show that
$\Lp$ and $\Lp'$ are bounded as $r\rightarrow r_{1}$.

Integrating \eqref{feq1} yields
\leqn{gbprop4.2}{
m(r) - m(r_{0}) = \int_{r_{0}}^{r} (NG+\rho^{-2}P)d\rho \geq
 \int_{r_{0}}^{r} NGd\rho
}
since $P \geq 0$ and $r_{0} > 0$. From \eqref{gbprop4.1}, \eqref{gbprop4.2},
and $N(r) \geq \epsilon $ it follows that
\eqn{gbprop4.3}{
2\epsilon\int^{r}_{r_{0}} G d\rho \leq 2
\int^{r}_{r_{0}} NG d\rho \leq r_{1} \qquad \forall r\in [r_{0},r_{1})\:,
}
which implies that
\leqn{gbprop4.4}{
2\int^{r}_{r_{0}}\rho^{-1} G d\rho 
\leq \frac{2}{r_{0}} \int_{r_{0}}^{r} G d\rho \leq
\frac{r_{1}}{\epsilon r_{0}}
\qquad \forall r\in [r_{0},r_{1}) \; .
}
Integrating \eqref{feq2} yields
\eqn{gbprop4.5}{
S(r) = S_{0} \exp(2\int^{r}_{r_{0}}\rho^{-1} G d\rho) \qquad (S_{0} > 0)\:,
}
and hence
\leqn{gbprop4.6}{
0 < S_{0} \leq S(r) \leq S_{0}\exp\left(\frac{r_{1}}{\epsilon r_{0}}\right)
\qquad \forall r\in [r_{0},r_{1}) \; .
}
by \eqref{gbprop4.4}.

Now,
\eqn{gbprop4.7}{
\norm{\Lp(r) - \Lp(r_{0}} = \norm{\int^{r}_{r_{0}}\Lp'(\rho)d\rho } \leq
\int^{r}_{r_{0}}\norm{\Lp'(\rho)}d\rho \; .
}
But,
\alin{gbprop4.8}{
\int^{r}_{r_{0}} \norm{\Lp'(\rho)}d\rho & \leq 
\left( \int^{r}_{r_{0}} \norm{\Lp'(\rho)}^{2}d\rho\right)^{\frac{1}{2}}
\sqrt{r-r_{0}} && \text{by H\"{o}lders inequality} \\
& = \left(2\int_{r_{0}}^{r} G d\rho \right)^{\frac{1}{2}} \sqrt{r-r_{0}}
&& \text{by def. of $G$} \\
& \leq \left(\frac{r_{1} (r_{1}-r_{0})}{ \epsilon } \right)^{\frac{1}{2}}
\: .
}
The above two results show that
\leqn{gbprop4.9}{
\sup_{r\in [r_{0},r_{1})}\norm{\Lp(r)} < \infty \: .
}
We can rewrite \eqref{feq3} as
\leqn{gbprop4.10}{
(NS\Lp')' = -\frac{S\Fc}{r^{2}} \; .
}
So then
\lgath{gbprop4.10}{
\norm{N(r)S(r)\Lp'(r) - N(r_{0})S(r_{0})\Lp'(r_{0})} 
= \norm{\int^{r}_{r_{0}} (NS\Lp')'d\rho} \notag \\
= \norm{\int^{r}_{r_{0}} \frac{S\Fc}{\rho^{2}} d\rho }
\leq \int^{r}_{r_{0}} \frac{S\norm{\Fc}}{\rho^{2}} d\rho \: .
\label{gbprop4.10.2}
}
But
\leqn{gbprop4.11}{
\norm{\Fc} = \norm{\frac{i}{2}(\Lo - [\Lp,\Lm])}
\leq \frac{1}{2}(\norm{\Lo} + \norm{[\Lp,\Lm]})
\leq \frac{1}{2}\norm{\Lo} + k\norm{\Lp}^{2}
}
for some constant $k > 0$ since $(X,Y) \rightarrow [X,c(Y)]$
is a continuous bilinear map from $\g\times \g$ to $\g$. It follows
from \eqref{gbprop4.6}, \eqref{gbprop4.9}, \eqref{gbprop4.10.2}, 
and \eqref{gbprop4.11} that
\eqn{gbprop4.12}{
\sup_{r\in [r_{0},r_{1})}\norm{\Lp'(r)} < \infty \: .
}
\end{proof}

From this point onward, we will  assume that $\Lo$ satisfies
the coercive condition \eqref{cc}. This next theorem can be
use to generalize theorem 7 of \cite{eymg} to any $\Lo$
that satisfies $\eqref{cc}$.

\begin{prop} \label{gbprop5} \mnote{[gbprop5]}
If $\{\Lp(r),m(r)\}$ is a solution to \eqref{feq1} and \eqref{feq3} on
$[r_{0},r_{1})$ $(r_{0} > 0)$ with $N(r) > 0$, then $\norm{\Lp(r)}^{2}$ can
not achieve a local maximum 
in the region where $\norm{\Lp(r)}^{2} > \frac{1}{2}\norm{\Lo}^{2}$.
\end{prop}
\begin{proof}
Let $v(r) := \norm{\Lp(r)}^{2}$ and suppose $v(r)$ achieves a local maximum
at $r_{*}$. Then 
\eqn{gbprop5.1}{
v'(r_{*}) = 0 \AND v''(r_{*}) \leq 0 \: .
}
From \eqref{feq3}, it is not hard to show that $v(r)$ satisfies
\leqn{gbprop5.2}{
r^{2}N(r)v''(r) + \Phi(r) v'(r) + 2 v(r) - \norm{[\Lp(r),\Lm(r)]}^{2} = 
2 r^{2} N(r) \norm{\Lp'(r)}^{2} \:,
}
where
\leqn{gbprop5.3}{
\Phi(r) := 2(m(r)-r^{-1}P(r)) = r(1-N(r))-2r^{-1}P(r) \: .
}
It follows from the above equations that
$v(r_{*}) \geq \frac{1}{2}\norm{[\Lp(r_{*}),\Lm(r_{*})]}^{2}$
while \eqref{cc} implies that
$\frac{4}{\norm{\Lo}^{2}} v(r_{*})^{2} \leq \norm{[\Lp(r_{*}),\Lm(r_{*})]}^{2}$.
Therefore $v(r_{*}) \leq \frac{1}{2} \norm{\Lo}^{2}$ and the proof is complete.
\end{proof}

The next proposition is very similar to the previous one, however
its slightly different conclusion will be useful in proving the next result.

\begin{prop} \label{gbprop6} \mnote{[gbprop6]}
Suppose  $\{\Lp(r),m(r)\}$ is a solution to \eqref{feq1} and \eqref{feq3} on
$[r_{0},r_{1})$ $(r_{0} > 0)$ with $N(r) > 0$ and let $v(r)=\norm{\Lp(r)}^{2}$.
If $v(r_{0}) > \frac{1}{2} \norm{\Lo}^{2}$ and $v'(r_{0}) > 0$ then
$v(r) > \frac{1}{2} \norm{\Lo}^{2}$ and $v'(r) > 0$ for all $r \geq r_{0}$.
\end{prop}
\begin{proof}
Let $r_{1}$ be the first $r > r_{0}$ such that $v'(r)=0$. Then \eqref{gbprop5.2}
shows that $r_{1}^{2} N(r_{1}) v''(r_{1}) \geq \norm{[\Lp(r_{1}),\Lm(r_{1})]}^{2} - 
2 v(r_{1})$ while it follows from \eqref{cc} that 
$v(r_{1})^{2}$ $\leq \frac{\norm{\Lo}^{2}}{4}\norm{[\Lp(r_{1}),\Lm(r_{1})]}^{2}$. Therefore
\eqn{gbprop6.1}{
r_{1}^{2} N(r_{1}) v''(r_{1}) \geq \frac{4}{\norm{\Lo}^{2}} v(r_{1})^{2} -
2v(r_{1}) \: .
}
But $v(r_{1}) > \frac{1}{2} \norm{\Lo}^{2}$ implies that 
$\frac{4}{\norm{\Lo}^{2}} v(r_{1})^{2} - 2v(r_{1}) > 0$ and hence $v''(r_{1}) > 0$
since $N(r_{1}) > 0$ by assumption. This implies that $v'(r_{1}) = 0$ is impossible. 
\end{proof}

The next proposition is a generalization of proposition 2.2 of \cite{k5719}.
The key to the proof is the observation that the equation 
\eqref{gbprop5.2} governing $\norm{\Lp(r)}^{2}$ can be
analyzed in the region where $\norm{\Lp(r)}^{2} > \norm{\Lo}^{2}/2$ 
using the techniques developed in \cite{k5719} 
for $G=SU(2)$. It is remarkable that the $SU(2)$ proof
can be adapted to the general case with such ease.

\begin{prop} \label{gbprop7} \mnote{[gbprop7]}
Suppose  $\{\Lp(r),m(r)\}$ is a solution to \eqref{feq1} and \eqref{feq3} 
defined in a neighborhood of $r_{0}$  and let $v(r)=\norm{\Lp(r)}^{2}$.
If $0 < N(r_{0}) < 1$, $v(r_{0}) > \frac{1}{2}\norm{\Lo}^{2}$ and
$v'(r_{0}) > 0$ then there exists a $r_{1} > r_{0}$ with $N(r_{1})=0$
, $0 < N < 1$ on $[r_{0},r_{1})$, and $[r_{0},r_{1})$ is the maximal 
interval of existence.
\end{prop}
\begin{proof}
Assume that the solution is defined on $[r_{0},\infty)$ and $N(r) > 0$.
Then \eqref{gbprop5.2} and \eqref{cc} imply that 
\leqn{gbprop7.1}{
r^{2}N v'' + \Phi v' + 2v- \frac{4}{\norm{\Lo}^{2}}v^{2} \geq 0 \: .
}
Consider the differential equation
\lgath{gbprop7.2}{
r^{2} N \vt'' + r\vt' + 2\vt - \frac{4}{\norm{\Lo}^{2}}\vt^{2} = 0 \, ,
\label{gbprop7.2.1} \\
\vt(r_{0}) = v(r_{0}) \AND \vt'(r_{0})=v'(r_{0}) \, . \label{gbprop7.2.2}
}
\begin{lem} \label{gblem8} \mnote{[gblem8]}
$v'(r) > \vt'(r)$ for all $r > r_{0}$ and hence $v(r) > \vt(r)$ for
all $r > r_{0}$.
\end{lem}
\begin{proof}
First note that it follows from proposition \ref{gbprop6} that 
\leqn{gbprop7.3}{
v'(r) > 0 \AND v(r) > \frac{1}{2}\norm{\Lo}^{2} \quad \forall \: r\geq r_{0} \: .
}
Because $N(r_{0}) > 0$, $P(r_{0})\geq 0$, and $v'(r_{0}) > 0$, we get 
from \eqref{gbprop5.3}, \eqref{gbprop7.1}, \eqref{gbprop7.2.1}, and \eqref{gbprop7.2.2} that
\leqn{gbprop7.4}{
r^{2}_{0}N(r_{0})(v''(r_{0})-\vt''(r_{0})) \geq -(\Phi(r_{0})-r_{0})v'(r_{0}) > 0 \: .
}
Thus $v''(r_{0}) > \vt''(r_{0})$ and hence $v'(r) > \vt'(r)$ for $r>r_{0}$ with
$r$ near $r_{0}$. Suppose $r_{1}$ is the first $r > 0$ for which
$v'(r_{1})=\vt'(r_{1})$. Then it follows from \eqref{gbprop7.2.2}, 
\eqref{gbprop7.3}, and the fact that $v'(r)>\vt'(r)$ for all $r\in [r_{0},r_{1})$
that
\leqn{gbprop7.6}{
\frac{4}{\norm{\Lo}^{2}} v(r_{1})^{2}-2v(r_{1}) > 
\frac{4}{\norm{\Lo}^{2}} \vt(r_{1})^{2}-2\vt(r_{1}) \: .
}
since the function 
\leqn{gbprop7.5}{
k(x) = \frac{4}{\norm{\Lo}^{2}} x^{2} - 2x > 0
}
is monotonically increasing in the region 
$x > \frac{1}{2}\norm{\Lo}^{2}$.
Then 
\eqn{gbprop7.7}{
r^{2}_{1}N(r_{1})(v''(r_{1})-\vt''(r_{1})) \geq 
\frac{4}{\norm{\Lo}^{2}} v(r_{1})^{2}-2v(r_{1}) - 
\left(\frac{4}{\norm{\Lo}^{2}} \vt(r_{1})^{2}-2\vt(r_{1})\right) > 0 
} 
by \eqref{gbprop7.1}, \eqref{gbprop7.2.1}, \eqref{gbprop7.6}, $N(r_{1}) >0$,
and $P(r_{1})\geq 0$.
Therefore $v''(r_{1}) > \vt''(r_{1})$ and this implies that $v'(r_{1})=\vt'(r_{1})$
is impossible.
\end{proof} 
\begin{lem} \label{gblem8a} \mnote{[gblem8a]}
$\vt'(r) > 0$ and $\vt(r) > \frac{1}{2}{\norm{\Lo^{2}}}$ for $r> r_{0}$. 
\end{lem}
\begin{proof}
Proved in similar fashion as proposition \ref{gbprop6}.
\end{proof}

\begin{lem} \label{gblem9} \mnote{[gblem9]}
If $f(r) = r\vt' + 2\vt-\frac{4}{\norm{\Lo}^{2}}\vt^{2}$, then there exists
an $R > r_{0}$ such that $f(r) < 0$ for all $r\geq R$.
\end{lem}
\begin{proof}
Suppose $f(r_{1}) = 0$ for some $r_{1} > r_{0}$.
Differentiating $f$ yields $f' = r \vt''+ \left( 3- \frac{8}{\norm{\Lo}^{2}} \vt \right) \vt'$. 
Since $f(r_{1}) = 0$, the 
differential equation \eqref{gbprop7.2.1} shows that 
\eqn{gbprop7.8a}{
r_{1}^{2} N(r_{1}) \vt''(r_{1}) = 0
}
and hence
$\vt''(r_{1})=0$ as $N(r_{1})>0$. Thus
\eqn{gbprop7.8}{
f'(r_{1}) = \left( 3-\frac{8}{\norm{\Lo}^{2}} \vt(r_{1}) \right) \vt'(r_{1}) < 0
}
by lemma \ref{gblem8a}. This shows that $f$ can cross zero at most once. Thus $f$ is
either always positive for $r > r_{0}$ or there exist an $R>r_{0}$ such that $f(r) < 0$
for all $r \geq R$. Suppose $f(r) > 0$ for all $r > r_{0}$. Then
\eqn{gbprop7.8a1}{
r\vt' + 2\vt - \frac{4}{\norm{\Lo}^{2}}\vt^{2} > 0
}
or equivalently
\eqn{gbprop7.9}{
\frac{d\vt}{-2\vt + \frac{4}{\norm{\Lo}^{2}}\vt^{2}} > \frac{dr}{r} 
}
by lemma \ref{gblem8a}. But 
\eqn{gbprop7.10}{
\int_{\vt(r_{0})}^{\infty} \frac{d\vt}{-2\vt + \frac{4}{\norm{\Lo}^{2}}\vt^{2}}
< \infty
}
while
\eqn{gbprop7.11}{
\int_{r_{0}}^{\infty} \frac{dr}{r} = \infty
}
which is a contradiction.
\end{proof}

Consider the differential equation
\lgath{gbprop7.12}{
r^{2} \vbr'' + r\vbr' + 2\vbr - \frac{4}{\norm{\Lo}^{2}}\vbr^{2} = 0
\label{gbprop7.12.1} \\
\vbr(R) = \vt(R)  \AND \vbr(R) = \vt'(R)  \label{gbprop7.12.2}
}
where $R$ is defined in lemma \ref{gblem9}.

\begin{lem} \label{gblem10} \mnote{[gblem10]}
$\vt(r) > \vbr(r)$ and $\vt'(r) > \vbr'(r)$ for all $r > R$.
\end{lem}
\begin{proof}
From \eqref{gbprop7.2.1}, \eqref{gbprop7.12.1} and \eqref{gbprop7.12.2} we have
\leqn{gbprop7.13}{
\vt''(R)-\vbr''(R) = \frac{f(R)}{R^{2}}\left(1-\frac{1}{N(R)}\right) \: .
}
Since $0 < N(r_{0}) < 1$ and we are assuming that $N > 0$, it follows from
proposition \ref{gbprop3} that $ 0 < N < 1$. Therefore  \eqref{gbprop7.13} shows
that $\vt''(R) > \vbr''(R)$ and hence $\vt'(r) > \vbr'(r)$ for $r > R$ with $r$
near $R$. Suppose there exists a smallest $r_{1} > R$ for which $\vt(r)=\vbr(r)$. Using
similar arguments as in proving lemma \ref{gblem8}, it can be shown that
\leqn{gbprop7.13a}{
\text{ $\vbr'(r) > 0$ and $\vbr(r) > \frac{1}{2}\norm{\Lo}^{2}$ for all $r > R$.} 
}
Thus
\eqn{gbprop7.14}{
\vt(r_{1}) > \vbr(r_{1}) > 0 \AND \frac{4}{\norm{\Lo}^{2}}\vt(r_{1})^{2}-2\vt(r_{1}) 
> \frac{4}{\norm{\Lo}^{2}}\vbr(r_{1})^{2}-2\vbr(r_{1})
}
and hence using \eqref{gbprop7.2.1} and \eqref{gbprop7.12.1} we see that
\eqn{gbprop7.15}{
r^{2}_{1}N(r_{1})\vt''(r_{1}) - r^{2}_{1}N(r_{1})\vbr''(r_{1}) =
\frac{4}{\norm{\Lo}^{2}}\vt(r_{1})^{2}-2\vt(r_{1}) - \left(
\frac{4}{\norm{\Lo}^{2}}\vbr(r_{1})^{2}-2\vbr(r_{1})  \right) > 0 
}
which implies that
\eqn{gbprop7.16}{
\vt''(r_{1}) > N(r_{1})\vt''(r_{1}) > \vbr''(r_{1}) \: .
}
Therefore   $\vt'(r_{1})=\vbr'(r_{1})$ is impossible.
\end{proof}
\begin{lem} \label{gblem11} \mnote{[gblem11]}
There exists a $\bar{r} > R$ for which
$\lim_{r\rightarrow \bar{r}} \vbr(r) = \infty$ and $\lim_{r\rightarrow \bar{r}} \vbr'(r) = \infty$.
\end{lem}
\begin{proof}
Let $t=\ln(r)$. Then we can write \eqref{gbprop7.12.1} as
\leqn{gbprop7.17}{
\ddot{\vbr} + 2\vbr - \frac{4}{\norm{\Lo}}^{2}\vbr^{2} = 0 \qquad 
\dot{(\cdot)} := \frac{d(\cdot)}{dt} \: .
}
So $\ddot{\vbr} > 0$ by \eqref{gbprop7.13a} and
\eqref{gbprop7.17}. This implies that $\dot{\vbr}$ is increasing. If $T=\ln(R)$ then $\dot{\vbr}(T) > 0$ by \eqref{gbprop7.13a},
and thus $\dot{\vbr}(t) > \dot{\vbr}(T) > 0$ for $t > T$ as $\dot{\vbr}$ is increasing. It
follows that $\lim_{t\rightarrow \infty} \vbr(t) = \infty$ .

The differential equation \eqref{gbprop7.17} admits a first integral 
\leqn{gbprop7.19}{
H(t) = \frac{1}{2}\dot{\vbr}^{2} + \vbr^{2} - \frac{4}{3\norm{\Lo}^{2}}\vbr^{3} \: .
}
Therefore if we let $H_{0} = H(T)$, then
\eqn{gbprop7.20}{
\frac{1}{2} \dot{\vbr}(t)^{2} = H_{0} + \vbr(t)^{2}\left(
\frac{4}{3\norm{\Lo}^{2}}\vbr(t) - 1 \right) \: 
}
as $H$ is a constant of the motion.
Since $\vbr(t)$ is increasing and $\lim_{t\rightarrow \infty} \vbr(t) = \infty$,
there exists a $t_{1} > T$ such that 
\eqn{gbprop7.21}{
\frac{1}{2}\dot{\vbr}^{2} > \frac{2}{3\norm{\Lo}^{2}}\vbr^{3} \quad \forall \; t\geq t_{1}
}
As  $\vbr > 0$ and $\dot{\vbr} > 0$, the above expression is equivalent to
\eqn{gbprop7.22}{
\frac{\dot{\vbr}}{\vbr^{\frac{3}{2}}} > \frac{4}{3\norm{\Lo}^{2}} \quad \forall \; t\geq t_{1}
\: .
}
Integrating both sides yields
\eqn{gbprop7.23}{
-\frac{2}{\sqrt{\vbr(t_{2})}} + \frac{2}{\sqrt{\vbr(t_{1})}} \geq
\frac{4}{3\norm{\Lo}^{2}}(t_{2}-t_{1}) \:,
}
or equivalently 
\eqn{gbprop7.24}{
\sqrt{\vbr(t_{2})} \geq \frac{2}{\frac{2}{\sqrt{\vbr(t_{1})}} + 
\frac{4}{3\norm{\Lo}^{2}}(t_{1}-t_{2})} \: .
}
This shows that there exists a $\bar{t}$ such that $\lim_{t\rightarrow\bar{t}} \vbr(t) = 
\infty$.  But $r= e^{t}$, so if we let $\bar{r} = e^{\bar{t}}$ then it follows that
$\lim_{r\rightarrow\bar{r}} \vbr(r) = \infty$ and 
$\lim_{r\rightarrow\bar{r}} \vbr'(r) = \infty$. 
\end{proof}

The above lemmas show that there exist a $\tilde{r} \leq \bar{r}$ such that 
$\lim_{r\nearrow \tilde{r}} v'(r) = \infty$ which proves that $\Lp(r)$ or $\Lp'(r)$
becomes unbounded as $r\rightarrow \tilde{r}$. This contradicts the solution existing
on $[r_{0},\infty)$. In view of proposition \ref{gbprop4}, we must have 
$N(\tilde{r}) = 0$. Let $r_{1}$ be the smallest $r$ such that
$N=0$.  Then \eqref{feq1v}
implies that 
\leqn{gbprop7.25}{
r_{1}N'(r_{1})= 1-\frac{2}{r^{2}_{1}}P(r_{1}) \; 
}
while  while it follows from \eqref{gbprop5.2} and \eqref{cc}
\alin{gbprop7.25a}{
\left( r_{1} - \frac{2}{r_{1}}P(r_{1})  \right)v'(r_{1})& + 
2v(r_{1}) - \frac{4}{\norm{\Lo}^{2}}v(r_{1})^{2} \\
\geq &
\left(\frac{\|[\Lp(r_{1}),c(\Lp(r_{1})]\|}{\|\Lp(r_{1})\|^{4}} -
\frac{4}{\norm{\Lo}^{2}}\right) v(r_{1})^{2} \geq 0 \, .
}
But $v(r_{1}) > \norm{\Lo}^{2}/2$ implies that
$4v(r_{1})^{2}/\norm{\Lo}^{2}-2v(r_{1}) > 0$ and
therefore
\eqn{gbprop7.26}{
\left(r_{1}-\frac{2}{r_{1}}P(r_{1})\right) v'(r_{1}) > 0 \: .
}
Since $v'(r_{1}) > 0$ the above inequality implies that $ r_{1} > 2P(r_{1})/r_{1}$ and
hence $1-2 P(r_{1})/(r_{1}^{2}) > 0$. Combining this with \eqref{gbprop7.25} we
see that $N'(r_{1}) > 0$ which contradicts $N(r_{1})=0$. Therefore $[r_{0},r_{1})$
must be the maximal interval of existence thus proving proposition \ref{gbprop7}.
\end{proof}

\begin{thm} \label{gbthm12} \mnote{[gbthm12]}
If $\{\Lp(r),m(r)\}$ is a solution to \eqref{feq1} and \eqref{feq3} defined on $[r_{0},\infty)$
$(r_{0} > 0)$ 
and it satisfies $N > 0$, $\norm{\Lp(r_{0})} < \norm{\Lo}/\sqrt{2}$ and $N(r_{0})<1$
then $\norm{\Lp(r)} < \norm{\Lo}/\sqrt{2}$ for
all $r \geq r_{0}$.
\end{thm}
\begin{proof}
Since $N$ cannot cross 1 from below by proposition \ref{gbprop3}, we have 
$0 < N(r) < 1$ for all $r\geq r_{0}$. Let $v(r) = \norm{\Lp(r)}^{2}$ and suppose 
there exists a $r_{1} > r_{0}$ such that $v(r_{1}) > \norm{\Lo}^{2}/2$. Then
by the mean value theorem there exists a $r_{*} \in (r_{1},r_{2})$ such that
$v'(r_{*}) > 0$ and $v(r_{*}) > \norm{\Lo}^{2}/2$. The proof then follows 
from proposition \ref{gbprop7}.
\end{proof}

The next theorem,  which guarantees that the mass is bounded, 
is a generalization of theorem 2  from \cite{hka25} and the proof uses similar 
methods.

\begin{thm} \label{gbthm13} \mnote{[gbthm13]}
If $\{\Lp(r),m(r)\}$ is a solution to \eqref{feq1} and \eqref{feq3} defined
on $[r_{0},\infty)$ with
$N > 0$, $N(r_{0}) < 1$ and $\norm{\Lp(r_{0})} \leq \norm{\Lo}/\sqrt{2}$ then
\eqn{gbthm13.1}{
\int_{r_{0}}^{\infty} NG \, dr < \infty \: .
}
\end{thm}
\begin{proof}
It follows from theorem \ref{gbthm12} that
\leqn{gbthm13.2}{
\norm{\Lp(r)}^{2} \leq \frac{1}{2} \norm{\Lo}^{2} \quad \forall \: r\geq r_{0} \: .
}
Let $\{X_{k}\}_{k=1}^{n}$ be a orthogonal basis for $V_{2}$ with normalization
$\norm{X_{k}} = 1/\sqrt{2}$. Define
\eqn{gbthm13.3}{
 w_{k}(r) := \rip{X_{k}}{\Lp(r)} \; .
}
Then it follows from \eqref{feq3} that for any  $q$,
\leqn{gbthm13.4}{
q r^{q-1} N w_{k}' = (r^{q}N w_{k}')' + 2 r^{q-1}N G w_{k}' + r^{q-2}
\rip{X_{k}}{\Fc} \: .
}
\begin{lem} \label{gblem14a} \mnote{[gblem14a]}
If $w_{k}$ has a critical point $c \in [r_{0},\infty)$ then
\eqn{gbthm13.5}{
\int_{r_{0}}^{\infty} N|w_{k}'|^{2} dr < \infty \: .
}
\end{lem}
\begin{proof}
Let $\Cc$ denote the set of critical points of $w_{k}(r)$. Since $w_{k}(r)$ is
analytic by proposition \ref{gbprop4} the set $\Cc$ can have no limit points.
There are two cases to consider, either $\Cc$ is bounded or $\Cc$ is unbounded.
Note that $\Cc$ is not empty by assumption. 

If $\Cc$ is bounded, let  $\tilde{c} = \sup \Cc$. Then $w_{k}'$ must
be either greater than zero or less than zero for $r > \tilde{c}$. We first
assume that $w_{k}'(r) > 0$ for $r > \tilde{c}$. Then integrating \eqref{gbthm13.4}
with $q=0$ yields
\eqn{gbthm13.6}{
Nw_{k}'(r) = -\int_{\tilde{c}}^{r} \frac{2}{\rho} NGw_{k}' d\rho -
\int_{\tilde{c}}^{r} \rho^{-2} \rip{X_{k}}{\Fc} d\rho
\leq \frac{1}{\sqrt{2}}\int_{\tilde{c}}^{r} \rho^{-2} \norm{\Fc} d\rho \: . 
}  
But it is clear from \eqref{gbprop4.11} and \eqref{gbthm13.2} that there
exists a $\beta > 0$ such that $\norm{\Fc(r)} \leq \sqrt{2}\beta$ for
all $r \geq r_{0}$. Consequently, $Nw_{k}'(r) \leq \beta(\tilde{c}^{-1}-r^{-1})$
and using 
\eqn{gbthm13.6a}{
\sup_{r_{0}\leq \rho \leq r} 
\beta\left(\frac{1}{\tilde{c}}-\frac{1}{\rho}\right)
= \beta\left(\frac{1}{\tilde{c}}-\frac{1}{r}\right)
}
we get
\lalign{gbthm13.7}{
\int_{\tilde{c}}^{r} N & |w'_{k}|^{2} d\rho  \leq  \beta\left(
\frac{1}{\tilde{c}}-\frac{1}{r}\right)\int_{\tilde{c}}^{r} w'_{k}(\rho) d\rho  
\leq \beta\left(\frac{1}{\tilde{c}}-\frac{1}{r}\right)|w_{k}(r)-w_{k}(\tilde{c})|
\notag \\
& \leq \beta\left(\frac{1}{\tilde{c}}-\frac{1}{r}\right)|\rip{X_{k}}{\Lp(r)-
\Lp(\tilde{c})}| \leq \frac{\beta}{\sqrt{2}}
\left(\frac{1}{\tilde{c}}-\frac{1}{r}\right)\norm{\Lp(r)-
\Lp(\tilde{c})} \notag \\
& \leq \frac{\beta}{\sqrt{2}}\left(\frac{1}{\tilde{c}}\right)
\alpha \label{gbthm13.7.3}
}
for some $\alpha >0$ by \eqref{gbthm13.2}. Letting $r\rightarrow \infty$ in the
above expression shows that
\eqn{gbthm13.8}{
\int_{\tilde{c}}^{\infty} N |w'_{k}|^{2} dr \leq \frac{\alpha\beta}{\sqrt{2}\tilde{c}}
 < \infty \: .
}
Similar arguments show that the above inequality continues to hold if $w_{k} < 0$.

If $\Cc$ is unbounded, there there exists a sequence of critical points $\{c_{j}\}$
such that $c_{j} < c_{j+1}$, 
$[c_{1},\infty) = \bigcup_{j=1}^{\infty}[c_{j},c_{j+1}] \:$,
and $w_{k}$ does not change sign on $(c_{j},c_{j+1})$. From \eqref{gbthm13.7.3} we
see that 
\eqn{gbthm13.9}{
 \int_{c_{j}}^{c_{j+1}} N |w'_{k}|^{2} dr 
\leq \frac{\alpha\beta}{\sqrt{2}}\left(\frac{1}{c_{j}}-\frac{1}{c_{j+1}}\right)
}
and so
\eqn{gbthm13.10}{
 \int_{c_{1}}^{c_{j}} N |w'_{k}|^{2} dr 
\leq \frac{\alpha\beta}{\sqrt{2}}\left(\frac{1}{c_{1}}-\frac{1}{c_{j}}\right) \: .
}
Letting $j\rightarrow \infty$ gives 
\eqn{gbthm13.10a}{
 \int_{c_{1}}^{\infty} N |w'_{k}|^{2} dr   
\leq \frac{\alpha\beta}{\sqrt{2}c_{1}} < \infty \: .
}
\end{proof}

\begin{lem} \label{gblem14} \mnote{[gblem14]}
If $w_{k}'> 0$ or $w_{k}' < 0$ for $r > r_{0}$ then for any $q > 1$ and $r > r_{0}$
\eqn{gbthm13.11}{
r^{q}N(r)|w_{k}'(r)| \leq r_{0}^{q}N(r_{0})|w_{k}'(r_{0})| + \frac{2}{3}
\sqrt{q}(r^{q}-r_{0}^{q}) + \int_{r_{0}}^{r} \norm{\Fc} \rho^{q-2} d\rho \: .
}
\end{lem}
\begin{proof}
Using Young's inequality it is not difficult to verify that
\leqn{gbthm13.12}{
q\int_{r_{0}}^{r} |w_{k}'|N \rho^{q-1} d\rho \leq
2\int_{r_{0}}^{r} |w_{k}'|^{3}N \rho^{q-1}d\rho + \frac{2}{3}
\sqrt{q}(r^{q}-r_{0}^{q}) \: .
}
Assume $w_{k}' > 0$. Then integrating \eqref{gbthm13.4} yields
\alin{gbthm13.13}{
r^{q}N(r)w_{k}'(r)& = r^{q}_{0}N(r_{0})w_{k}'(r_{0}) + \\ 
& q\int_{r_{0}}^{r}
\left(\rho^{q-1}N w_{k}'- 2\rho^{q-1}N G w_{k}'\right)\, d\rho 
-\int_{r_{0}}^{r} \rho^{q-2}\rip{X_{k}}{\Fc} \, d\rho \, .
}
But
\eqn{gbthm13.14}{
|w_{k}'|^{2} = |\rip{\Lp'}{X}|^{2} \leq \frac{1}{2}\norm{\Lp'}^{2} = G
\AND |\rip{X_{k}}{\Fc}| \leq \norm{X_{k}}\norm{\Fc} \leq \norm{\Fc}
} and therefore
\alin{gbthm13.15}{
r^{q}N(r)w_{k}'(r) & 
\leq  r^{q}_{0}N(r_{0})w_{k}'(r_{0}) + \\ 
q\int_{r_{0}}^{r}&
\left(\rho^{q-1}N w_{k}'- 2\rho^{q-1}N|w_{k}'|^{2}w_{k}'\right)\, d\rho 
+ \int_{r_{0}}^{r} \norm{\Fc} \rho^{q-2} d\rho\\ 
& \leq r_{0}^{q}N(r_{0})|w_{k}'(r_{0})| + \frac{2}{3}
\sqrt{q}(r^{q}-r_{0}^{q}) + \int_{r_{0}}^{r} \norm{\Fc} \rho^{q-2} d\rho 
&& \text{by \eqref{gbthm13.12}}
}
Similar arguments show that if $w_{k}' < 0$ then
\eqn{gbthm13.16}{
-r^{q}N(r)w_{k}'(r) \leq -r_{0}^{q}N(r_{0})|w_{k}'(r_{0})| + \frac{2}{3}
\sqrt{q}(r^{q}-r_{0}^{q}) + \int_{r_{0}}^{r} \norm{\Fc} \rho^{q-2} d\rho \: .
}
\end{proof}

\begin{lem} \label{gblem15} \mnote{[gblem15]}
If $w_{k}' >0$ or $w_{k}' < 0$  for $r > r_{0}$ then there exists a constant $h > 0$ such that
$N|w_{k}'| \leq h$ for $r > r_{0}$.
\end{lem}
\begin{proof}
Now, $\norm{\Fc} = \|[\Fh,\Lp]\| \leq h_{1}\|\Fh\|$ for some constant
$h_{1} > 0$ since $[\cdot,\cdot]$ is a continuous bilinear map from $\g\times \g$
to $\g$. So then
\eqn{gbthm13.17}{
N(r)|w_{k}'(r)| \leq 
\left(\frac{r_{0}}{r}\right)^{q}N(r_{0})|w_{k}'(r_{0})| + \frac{2}{3}
\sqrt{q}\left(1-\left(\frac{r_{0}}{r}\right)^{q}\right) + 
\frac{h_{1}}{r^{q}}\int_{r_{0}}^{r} \norm{\Fc} \rho^{q-2} d\rho \: .
}
by lemma \ref{gblem14}. But 
\alin{gbthm13.18}{
\int_{r_{0}}^{r} \norm{\Fc} \rho^{q-2} d\rho & \leq
\left(\int_{r_{0}}^{r} \norm{\Fc}^{2} \rho^{-2} d\rho \right)^{\frac{1}{2}}
\left(\int_{r_{0}}^{r} \rho^{2(q-1)} d\rho \right)^{\frac{1}{2}} \\
& = \left(\int_{r_{0}}^{r} \norm{\Fc}^{2} \rho^{-2} d\rho \right)^{\frac{1}{2}}
\sqrt{\frac{r^{2q-1}-r_{0}^{2q-1}}{2q-1}} \: .
}
Combining the above two inequalities yields
\lalign{gbthm13.19}{
N(r)|w_{k}'(r)| & \leq 
\left(\frac{r_{0}}{r}\right)^{q}N(r_{0})|w_{k}'(r_{0})| + \frac{2}{3}
\sqrt{q}\left(1-\left(\frac{r_{0}}{r}\right)^{q}\right) + \notag \\
& \frac{h_{1}}{\sqrt{2q-1}}\left(\int_{r_{0}}^{r} \norm{\Fc}^{2} 
\rho^{-2} d\rho \right)^{\frac{1}{2}}\sqrt{\frac{r^{2q-1}-r_{0}^{2q-1}}{r^{2q}} } \: .
\label{gbthm13.19.2}
}
From \eqref{feq1} we have $r^{_2}P = r^{-2}\|\Fh\|^{2}/2 \leq m'$, while
$N > 0$ implies $2m(r) < r$  and hence 
\leqn{gbthm13.20}{
\int_{r_{0}}^{r} \norm{\Fc}^{2}\rho^{-2} d\rho  \leq 2m(r) - 2m(r_{0})
 \leq r \: .
}
So
\eqn{gbthm13.21}{
N(r)|w_{k}'(r)|  \leq
\left(\frac{r_{0}}{r}\right)^{q}N(r_{0})|w_{k}'(r_{0})| + \frac{2}{3}
\sqrt{q}\left(1-\left(\frac{r_{0}}{r}\right)^{q}\right) + 
 \frac{h_{1}}{\sqrt{2q-1}}\sqrt{1-\left(\frac{r_{0}}{r}\right)^{2q-1} } 
}
by \eqref{gbthm13.19.2} and \eqref{gbthm13.20}. Setting $q=2$ in the above
expression yields
\eqn{gbthm13.22}{
N(r)|w_{k}'(r)|  \leq N(r_{0})|w_{k}'(r_{0})| + \frac{2}{3}
\sqrt{2}+ \frac{h_{1}}{\sqrt{3}} \: .
}
\end{proof}

\begin{lem} \label{gblem16} \mnote{[gblem16]}
If $w_{k}' >0$ or $w_{k}' < 0 $  for $r > r_{0}$ then 
\eqn{gthm13.23}{
\int_{r_{0}}^{\infty} N |w_{k}'|^{2} dr < \infty \: .
}
\end{lem}
\begin{proof}
Suppose $w_{k}' > 0$. Then
\alin{gbthm13.24}{
\int_{r_{0}}^{r} N |w_{k}'|^{2} d\rho & \leq h\int_{r_{0}}^{r} w_{k}'\, d\rho
&& \text{by lemma \ref{gblem15} } \\
& \leq h|w_{k}(r)-w_{k}(r_{0})| = h |\rip{X_{k}}{\Lp(r)-\Lp(r_{0})}| \\
& \leq \frac{h}{\sqrt{2}}( \norm{\Lp(r)} + \norm{\Lp(r_{0})})  \leq K
}
for some constant $K > 0$ by \eqref{gbthm13.2}. Letting $r \rightarrow \infty$
in the above expression completes the proof. 
\end{proof}

Now,
\eqn{gbthm13.25}{
G = \frac{1}{2}\norm{\Lp'}^{2} = \frac{1}{2}
\sum_{k=1}^{n}|\rip{\Lp'}{\norm{X_{k}}^{-1}X_{k}}| = \sum_{k=1}^{n} |w_{k}'|^{2} \: .
}
Therefore
\eqn{gbthm13.25a}{
\int_{r_{0}}^{\infty} NG\, dr = \sum_{k=1}^{n} \int_{r_{0}}^{\infty} N|w_{k}'|^{2} dr
< \infty
}
by lemmas \ref{gblem14a} and \ref{gblem16}.
\end{proof}

\begin{cor} \label{gbcor17} \mnote{[gbcor17]}
If $\{\Lp(r),m(r)\}$ is a solution to \eqref{feq1} and \eqref{feq3} defined on
$[r_{0},\infty)$ with
$N > 0$, $N(r_{0}) < 1$ and $\norm{\Lp(r_{0})} \leq \norm{\Lo}/2$ then
$\lim_{r\rightarrow \infty} m(r)$ exists and $\lim_{r\rightarrow \infty} N(r) = 1$
\end{cor}
\begin{proof}
Since $N > 0$, $P\geq 0$,$G \geq 0$, and $\norm{\Lp(r)}\leq \norm{\Lo}/2$ it
follows from \eqref{feq1} that
\eqn{gbcor17.1}{
0 \leq m' \leq NG + \frac{K}{r^{2}}
}
for some constant $K > 0$. Integrating yields 
\eqn{gbcor17.2}{
m(r) \leq m(r_{0}) + \int_{r_{0}}^{\infty} NG\, dr + \frac{K}{r_{0}} < \infty
}
by theorem \ref{gbthm13}. Thus $\lim_{r\rightarrow \infty} m(r)$ exists as 
$m$ is increasing and bounded above. From the definition of $N$ it is then
clear that $\lim_{r\rightarrow \infty} N(r) = 1$.
\end{proof}

\begin{prop} \label{gbprop18} \mnote{[gbprop18]}
If $\{\Lp(r),m(r)\}$ is a solution to \eqref{feq1} and \eqref{feq3} defined
on $[r_{0},\infty)$ with
$N > 0$, $N(r_{0}) < 1$ and $\norm{\Lp(r_{0})} \leq \norm{\Lo}/2$ then
\eqref{feq2} can be solved for $S$ and $S(r_{0})$ can be chosen so that
$\lim_{r\rightarrow \infty} S(r) = 1$.
\end{prop}
\begin{proof}
We can solve equation \eqref{feq2} to get
$S(r) = S_{0}\exp(\int_{r_{0}}^{r} 2\rho^{-1}G\, d\rho)$,
where $S_{0} > 0$ is an arbitrary constant.
Because $N > 0$ and $\lim_{r\rightarrow \infty} N(r) = 1$ by corollary \ref{gbcor17},
$N$ is bounded below on $[r_{0},\infty)$ by a positive constant $\hat{N}$. Then
\eqn{gbprop18.1}{
\int_{r_{0}}^{\infty} \frac{2}{r}G\, dr = \int_{r_{0}}^{\infty} \frac{2}{Nr}NG\, dr
\leq \frac{2}{r_{0}\hat{N}} \int_{r_{0}}^{\infty} NG\, dr < \infty
}
by theorem \ref{gbthm13}. So we can let $S_{0} = \exp(-\int_{r_{0}}^{\infty} 2r^{-1}G\, dr)$
which then implies that $\lim_{r\rightarrow \infty} S(r) = 1$.
\end{proof}

\begin{prop} \label{gbprop19} \mnote{[gbprop19]}
If $\{\Lp(r),m(r)\}$ is a solution to \eqref{feq1} and \eqref{feq3} defined
on $[r_{0},\infty)$ with
$N > 0$, $N(r_{0}) < 1$ and $\norm{\Lp(r_{0})} \leq \norm{\Lo}/2$ then
there exists a constant $h > 0$ such that $rN\norm{\Lp'}^{2} < h$ for
all $r \geq r_{0}$.
\end{prop}
\begin{proof}
From corollary \ref{gbcor17} and theorem  \ref{gbthm12}, we get that 
$P(r)$ is bounded and $\lim_{r\rightarrow \infty} m(r)=m_{\infty}$ for some constant
$m_{\infty} > 0$. Then from  the definition of $\Phi(r)$ (see \eqref{gbprop5.3}) it
is clear that there exists a $r_{*}$ and an $\epsilon > 0$ such that 
$\Phi(r) \geq \epsilon > 0$ for all $r\geq r_{*}$. Thus
\leqn{gbprop19.1}{
\frac{\Phi(r)}{r^{2}N} + \frac{2G}{r} > 0 \quad \forall \: r>r_{*} \: .
}
Because $N > 0$ and $\lim_{r\rightarrow \infty} N(r) = 1$ by corollary \ref{gbcor17},
$N$ is bounded below on $[r_{0},\infty)$ by a positive constant $\hat{N}$. Also
note that theorem \ref{gbthm12} and \eqref{gbprop4.11} imply that $\norm{\Fc}$ is
bounded.  
So then 
\lalign{gbprop19.2z}{
\int_{r_{*}}^{\infty} \frac{1}{r} |\rip{\Lp'}{\Fc}|\, dr
 \leq & \int_{r_{*}}^{\infty} \sqrt{N}\norm{\Lp'}\frac{1}{\sqrt{N}r}\norm{\Fc}\, dr \notag \\
 \leq & \frac{1}{2} \int_{r_{*}}^{\infty}\left(NG + \frac{1}{Nr^{2}}\norm{\Fc}^{2}\right) \, dr
< \infty  \label{gbprop19.2}
}
by theorem \ref{gbthm13}.

From \eqref{feq3} and \eqref{feq1v} it follows that 
\eqn{gbprop19.3}{
(rNG)' = -\left(\frac{\Phi(r)}{r^{2}N} + \frac{2G}{r}\right)rNG +NG -
\frac{1}{r}\rip{\Lp'}{\Fc} \: .
}
Therefore
\eqn{gbprop19.4}{
rN(r)G(r) = e^{-\Psi(r)}\left( rN(r_{*})G(r_{*}) + \int_{r_{*}}^{r}\left(
NG - \frac{1}{\rho}\rip{\Lp'}{\Fc}\right)e^{\Psi(\rho)}\,d\rho\right) \:,
}
where $\Psi(r) = \int_{r_{*}}^{r}(\rho^{-2}N^{-1}\Phi+2\rho^{-1}G)\, d\rho$.
It then follows from theorem 
\ref{gbthm13}, \eqref{gbprop19.1}, \eqref{gbprop19.2}, and
\eqn{gbprop19.5}{
\sup_{r_{*}\leq\rho\leq r} \exp(\Psi(\rho)) = \Psi(r)
}
that there exists
a constant $\tilde{h}$ such that $rN(r)G(r) < \tilde{h}$ for all $r \geq r_{*}$.
Letting 
\eqn{gbprop19.6}{
h = \max\{\,\tilde{h}\,, \,
\max\{\,\rho N(\rho)G(\rho)\,|\, r_{0}\leq \rho \leq r_{*}\,\}\,\}
}
completes the proof.
\end{proof}

\begin{prop} \label{gbprop20} \mnote{[gbprop20]}
If $\{\Lp(r),m(r)\}$ is a solution to \eqref{feq1} 
and \eqref{feq3} defined
on $[r_{0},\infty)$ with
$N > 0$, $N(r_{0}) < 1$ and $\norm{\Lp(r_{0})} \leq \norm{\Lo}/2$ then
there exist a constant $h > 0$ such that
$r\norm{\Lp'} \leq h$ for all $r\geq r_{0}$.
\end{prop}
\begin{proof}
From theorem \ref{gbthm12} and \eqref{gbprop4.11} we see that
\eqn{gbrop20.2}{
\sup_{r_{0}\leq r< \infty} S(r)\norm{\Fc} < K
}
for some constant $K > 0$.
So then
\leqn{gbprop20.3}{
\norm{N(r)S(r)\Lp'(r)-N(r_{1})S(r_{1})\Lp'(r_{1})}
\leq \frac{K}{r_{0}} \quad \forall r,r_{1} \: \in [r_{0},\infty)
}
by \eqref{gbprop4.10}. 
An immediate consequence of corollary \ref{gbcor17} and 
propositions \ref{gbprop18} and \ref{gbprop19} is that
\eqn{gbprop20.1}{
\lim_{r\rightarrow\infty} N(r)S(r)\Lp'(r) = 0 \: .
}
So then letting $r_{1} \rightarrow \infty$ in \eqref{gbprop20.3} yields
$\norm{N(r)S(r)\Lp'(r)} \leq K/r_{0}$ for all $r \geq r_{0}$. 
However, we know from corollary \ref{gbcor17} and proposition \ref{gbprop18} 
that both $S(r)$ and $N(r)$ 
are bounded below by a positive number  $\epsilon > 0$. Therefore if we let
$h = K/(r_{0}\epsilon^{2})$ then 
$\norm{\Lp'(r)} \leq h$ for all $r \geq r_{0}$.
\end{proof}

We are now ready to prove theorem \ref{gbthm21}.
\begin{proof}[Proof of theorem \ref{gbthm21}]
{\em (i)}-{\em (v)} : These are just a restatement of 
corollary \ref{gbcor17}, theorem \ref{gbthm12}, and 
propositions \ref{gbprop1}, and \ref{gbprop3}.
\\
{\em (vi)} : Since $N$ is bounded below away from zero and $N \rightarrow 1$ as 
$r \rightarrow \infty$, the 
change of variable from $r$ to $\tau$ given by \eqref{tau} 
is well defined and $\tau \rightarrow \infty$ 
as $r \rightarrow \infty$. 
Therefore to prove {\em (vi)} we can show instead that
$\norm{\Lp(\tau)-\Ff^{\times}} \rightarrow 0$ and $\dot{\Lambda}_{+}(\tau) \rightarrow 0$
as $\tau \rightarrow \infty$. Using \eqref{feqBFM1}-\eqref{feqBFM6} it is easy to
show that $\Lp(\tau)$ satisfies
\leqn{gbthm21.1}{
\ddot{\Lambda}_{+} - \dot{\Lambda}_{+} + \Fc = \delta(\tau)\dot{\Lambda}_{+} \;,
}
where
\leqn{gbthm21.2}{
\delta(\tau) = 2\mu(\tau) - \kappa(\tau) -1 \: .
}
Since $\lim_{r \rightarrow \infty} N(r) = 1$ we have 
$\lim_{\tau \rightarrow \infty} \mu(\tau) = 1$ and hence 
it follows from proposition \ref{gbprop20}
that $\lim_{\tau \rightarrow \infty} \Lp'(\tau) = 0$. Therefore
$\lim_{\tau \rightarrow \infty}  \mu(\tau)^{2}G(\tau)  = 0$.
Also, because $\Lp(r)$ is bounded and hence $\Lp(\tau)$ is also bounded, we get
$\lim_{\tau \rightarrow \infty} r(\tau)^{-2}P(\tau) = 0$.
From the definition \eqref{BFMvars} of $\kappa$ it is then clear that
$\lim_{\tau \rightarrow \infty} \kappa(\tau) = 1$ and hence 
\eqn{gbthm21.3}{
\lim_{\tau \rightarrow \infty} \delta(\tau) = 0 \; .
}
Another consequence of proposition 
\ref{gbprop20} is that $\dot{\Lambda}_{+}(\tau)$ is bounded. Therefore we
see that
$\mathbf{X}(\tau) = (\Lp(\tau),\dot{\Lambda}_{+}(\tau))$ is a bounded, non-trivial
solution to the differential equation \eqref{afym1}. So $\norm{\Lp(\tau)-\Ff^{\times}} 
\rightarrow 0$ and $\dot{\Lambda}_{+}(\tau) \rightarrow 0$ 
as $\tau \rightarrow \infty$ by theorem 
\ref{aymthm5}.

If $S_{\lambda}$ is a $\Pi$-system then it follows from the
discussion in section \ref{regA1} that
\eqn{gbthm21.3a}{
\Lp(r) = \sum_{\alpha \in S_{\lambda}} w_{\alpha}(r) \eb_{\alpha} 
}
where the $w_{\alpha}(r)$ are real valued functions.
Therefore $\Lp(\tau) \in E_{+}$ for all $\tau$.
Since $\Ff^{\times}\cap E_{+}$ is a discrete set by
lemma \ref{coerlem1a} and
$\norm{\Lp(\tau)-\Ff^{\times}}
\rightarrow 0$ as $\tau \rightarrow \infty$, there exists
a $\Op^{\infty} \in \Ff^{\times}\cap E_{+}$ such that
$\lim_{r\rightarrow \infty} \Lp(r) = \Op^{\infty}$.
\end{proof}

We now show that any local solution that can
be continued out to a global solution necessarily
satisfies \eqref{gbidata1} and \eqref{gbidata2}.
\begin{prop} \label{gbprop22} \mnote{[gbprop22]}
Suppose $\{\Lp(r),m(r)\}$ is a local solution to \eqref{feq1}
and \eqref{feq3} defined in a neighborhood of
$r=r_{*}$ where $r_{*}=0$ or $r_{*}=r_{H}>0$.
If the local solution can be continued out to
$r=\infty$ with $N(r) > 0$ for $r>r_{*}$, then
\eqn{gbidataB}{
N(r_{0}) < 1 \, , \quad \norm{\Lp(r_{0})} \leq \frac{1}{\sqrt{2}}\norm{\Lo}\, , \quad
[\Lp'(r_{0}),\Lm(r_{0})] + [\Lm'(r_{0}),\Lp(r_{0})] = 0 \, ,
} for some $r_{0} > r_{*}$.
\end{prop}
\begin{proof}
$\mathbf{r_{*}=0}$: We know by
theorem 2 in \cite{eymgg} that $m(r)$ and $\Lp(r)$
are analytic in a neighborhood of $r=0$
and
\eqn{gbprop22.1}{
m(r) = \text{O}(r^{3}) \AND \Lp(r) = \Op + X r^{2} + \text{O}(r^{3})
\quad \text{as $r\rightarrow 0$}
}for some $X\in E_{1}$. Substituting powerseries
representation for $m(r)$ and $\Lp(r)$ into the
field equations \eqref{feq1}and \eqref{feq3}
shows that $m(r) = \|X\|^{2}r^{3} + \text{O}(r^{4})$ near $r=0$ and
hence
\leqn{gbprop22.2}{
\text{$N(r) = 1- 2\|X\|^{2}r^{2} + \text{0}(r^{3})$ as 
$r\rightarrow 0$.}
}
Let $v(r) = \|\Lp(r)\|^{2}$. Then
\alin{gbprop22.3}{
v(0) & = \half \rip{[\Lo,\Op]}{\Op} = \half \rip{\Lo}{[\Op,\Om]}
&& \text{by \eqref{rip} and $\Op = -c(\Om)$} \\
& = \half \|\Lo\|^{2} && \text{since $[\Op,\Om] = \Lo$}
}

The  solution is trivial for $X=0$ by proposition 
\ref{gbprop2}. If we assume the solution is non-trivial,
then we must have $X \neq 0$. 
So \eqref{gbprop22.2} shows there exists an $\epsilon > 0$ such
that $0 < N(r) < 1$ for $0 < r < \epsilon$. 
Suppose  there exists a $r_{0} \in (0,\epsilon)$
for which $v(r_{0}) > \|\Lo\|^{2}/2$. By the mean value theorem
there exist a $r_{1}\in (0,r_{0})$ such that
$v(r_{1}) > \|\Lo\|^{2}/2$ and $v'(r_{1}) > 0$. It then follows
from proposition \ref{gbprop7} that the maximal interval
of existence for the solutions is finite which contradicts
the assumption that it is defined on $(0,\infty)$.
Therefore we conclude that $v_{r} \leq \|\Lo\|^{2}/2$
for all $0 < r < \epsilon$.
To complete the proof for $r_{*}=0$. we observe that
$[\Lp'(r),\Lm(r)] +[\Lm'(r),\Lp(r)]=0$ near 
$r=0$ by theorem 3 of \cite{eymgg}.
${}$
\vspace{0.2cm}\\
$\mathbf{r_{*}=r_{H}}$: From the boundary conditions at
$r=r_{H} > 0$ we have $N(r_{H})=0$ and $N'(r_{H}) > 0$
and hence $0 < N(r) < 1$ for $r>r_{H}$ with $r$ near $r_{H}$.
We know by theorem 7 in \cite{eymgg}  that $\Lp(r)$ and $N(r)$
are analytic in the variable $t=r-r_{H}$ near $t=0$ 
and there exists a $X\in V_{2}$ so that
\eqn{gbprop22.4}{
N(t) = \nu t + \text{0}(t^{2})  \AND
\Lp(t) = X + \text{0}(t) \, ,
}
where
\eqn{gbprop22.5}{
\nu = \frac{1}{\rh}-\frac{1}{\rh^{3}}\norm{\Lo+[X,c(X)]}^{2} > 0 \, .
}
Expanding  $N(t)$ and $\Lp(t)$ in powerseries about $t=0$, it
follows from the field equations \eqref{feq1} and \eqref{feq3}
that
\leqn{gbprop22.6}{
\Lp'(r_{H}) = \frac{1}{2\nu r_{H}^{2}}[X,\Lo + [X,c(X)]] \; .
}
Note also that
 $[\Lp'(r),\Lm(r)] +[\Lm'(r),\Lp(r)]=0$ for $r$ near
$r_{H}$ by theorem 8 in \cite{eymgg}.

Let $v(r)=\|\Lp(r)\|^{2}$. If $v(r_{H}) < \|\Lo\|^{2}/2$
then $v(r) < \|\Lo\|^{2}/2$ for $r$ near $r_{H}$ and
we are done. So assume that $v(r_{H})\geq \|\Lo\|^{2}/2$.
Now,
\eqn{gbprop22.7}{
v'(r_{H}) = 2\rip{\Lp'(r_{H})}{\Lp(r_{H})}
= \frac{1}{\nu r_{H}^{2}}\rip{[X,[\Lo + [X,c(X)]]}{X} \; 
}
by \eqref{gbprop22.6}.  Using \eqref{rip}and $X\in V_{2}$, we 
can write the above expression as
\leqn{gbprop22.8}{
v'(r_{H}) = \frac{1}{\nu r_{H}^{2}}(\norm{[X,c(X)]}^{2}-2
\norm{X}^{2})\; .
}
But
\leqn{gbprop22.9}{
0 \leq \norm{\Lo+[X,c(X)]}^{2} =
(\norm{\Lo}^{2}-2\norm{X}^{2})  + 
(\norm{[X,c(X)]}^{2}-2\norm{X}^{2}) 
}
by \eqref{rip} and the fact that $X \in V_{2}$.
Since $X=\Lp(r_{H})$,
\leqn{gbprop22.10}{
v'(r_{H}) \leq 0 \Longleftrightarrow \|\Lp(r_{H})\|^{2} \leq
\frac{1}{2}\|\Lo\|^{2}
}
by \eqref{gbprop22.8} and \eqref{gbprop22.9}. 
Suppose $v(r_{H})> \|\Lo\|^{2}/2$. Then $v'(r_{H}) > 0$
by \eqref{gbprop22.10}.
But this implies that the maximum interval of existence
for the solutions is finite by proposition \ref{gbprop7}.
So $v(r_{H})=\|\Lo\|^{2}/2$ as the
solution is assumed to exist on $(r_{H},\infty)$.
Suppose now there exists a $r_{0} > r_{H}$ with $r_{0}$ near $r_{H}$
for which $v(r_{0}) > \|\Lo\|^{2}/2$. Then by the mean value
theorem there exist a $r_{1} \in (r_{H},r_{0})$ so
that $v(r_{1}) > \|\Lo\|^{2}/2$ and $v'(r_{1}) > 0$.
But this is impossible by proposition \ref{gbprop7}. Therefore
we must have $v(r) \leq \|\Lo\|^{2}/2$ for $r > r_{H}$ and
$r$ near $r_{H}$. 
\end{proof}

\bigskip

\noindent \emph{Acknowledgements}. We would like to thank M. Li. and
M. Nevins for useful discussions and advice and also R. Bartnik and L. Kramer
for helpful suggestions in regards to the proof of lemma \ref{rA1lem}. This
work was partially supported by the NSERC grant A8059 at the University of
Alberta and the ARC grant A00105048 at the University of Canberra.

\end{document}